\documentclass[aps, pra, twocolumn, longbibliography, floatfix, nobibnotes, 10pt, superscriptaddress]{revtex4-2}

\usepackage{color}
\usepackage{dcolumn}
\usepackage{graphicx}

\usepackage{amsmath,amsfonts,amssymb,amsthm}
\usepackage{mathtools}

\usepackage{bbold}
\usepackage{dsfont}

\usepackage{soul}

\usepackage{fancyhdr}
\usepackage{multirow}
\usepackage{bm}
\usepackage{dsfont}
\usepackage{cancel}

\usepackage{xifthen}

\usepackage{float}
\graphicspath{{"./graphs/"}{"./scheme/"}}

\usepackage{graphicx}
\usepackage[dvipsnames,dvipsnames,svgnames]{xcolor}
\usepackage[colorlinks,bookmarks=false,
linkcolor=blue,
urlcolor=blue,
citecolor=blue
]{hyperref}

\newcommand{\appref}[1]{Appendix~\ref{#1}}

\usepackage{xspace}

\renewcommand{\Re}{\text{Re}}

\newcommand{\q}[1]{``#1"}
\usepackage{physics}

\usepackage{xifthen}

\def\op#1{\hat{#1}}
\renewcommand{\ao}[1][]{%
	\ifthenelse{\equal{#1}{}}{\ensuremath{\op{a}}}{\ensuremath{\op{#1}}}%
}

\newcommand{\co}[1][]{%
	\ifthenelse{\equal{#1}{}}{\ensuremath{{{}\op{a}^{\dagger}}}}{\ensuremath{{{}\op{#1}^{\dagger}}}}%
}

\makeatletter
\newcommand*{\transpose}{\bgroup\@transpose}
\newcommand*{\@transpose}[1][0]{\mathpalette\@@transpose{#1}\egroup}
\newcommand*{\@@transpose}[2]{\setbox0=\hbox{\m@th$#1\mkern-#2mu\intercal$}\raise\dp0\box0}
\makeatother

\usepackage{enumitem}
\usepackage{booktabs}

\usepackage{array,calc}

\usepackage{pgfplots} 
\pgfplotsset{compat=1.18}
\usetikzlibrary{plotmarks}

\usepackage{scalerel}

\usepackage[normalem]{ulem}

\graphicspath{{"figures/"}}

\newcommand{\LC}[1]{\ensuremath{\mathrm{LC}_{#1}}}

\usepackage{placeins}

\begin{document}
	
	\title{Phase-resolved multichannel quantum escape between limit cycles}
	%
	\author{Caroline Nowoczyn}
	\affiliation{Center for Optical Quantum Technologies and Institute for Quantum Physics, University of Hamburg, Hamburg 22761, Germany}
	\affiliation{The Hamburg Center for Ultrafast Imaging, Hamburg 22761, Germany}
	\author{Ludwig Mathey}
	\affiliation{Center for Optical Quantum Technologies and Institute for Quantum Physics, University of Hamburg, Hamburg 22761, Germany}
	\affiliation{The Hamburg Center for Ultrafast Imaging, Hamburg 22761, Germany}
	\author{Kilian Seibold}
	\affiliation{Department of Physics, University of Konstanz, 78464 Konstanz, Germany}
	\date{\today}

\begin{abstract}
Driven-dissipative quantum systems can recover stable dynamical attractors in the semiclassical limit, including coexisting limit cycles.
At finite fluctuation strength, this classical coexistence becomes quantum metastability: the corresponding oscillatory states undergo rare fluctuation-induced transitions.
We demonstrate phase-resolved quantum escape between two such states in a driven optomechanical resonator.
Unlike escape from fixed points, switching between extended attractors occurs across a periodic basin boundary and depends on the phase at which fluctuations approach it.
Using quantum-jump trajectories across a controlled quantum-to-classical crossover, we reconstruct the escape geometry directly from switching events.
Escape from the small-amplitude cycle proceeds through a single radial corridor and exhibits near-Arrhenius scaling, whereas escape from the large-amplitude cycle involves competing phase-localized corridors with distinct effective activation scales.
The resulting curvature in the switching-rate scaling, together with event-conditioned phase distributions, identifies finite-fluctuation multichannel quantum escape between extended attractors.
\end{abstract}

\maketitle
\section{Introduction}
\label{sec:intro}

Metastability in driven-dissipative quantum systems is characterized by long-lived states and rare transitions between them~\cite{Macieszczak2016,Macieszczak2021}.
These states often arise as quantum counterparts of coexisting semiclassical attractors, such as fixed points~\cite{Andersen2020,Marthaler2006,Carde2026} or limit cycles~\cite{Loerch2014,Gao2015,BenArosh2021}.
Quantum extensions of semiclassical limit cycles have been realized in optomechanical, superconducting, electromechanical, and trapped-ion platforms~\cite{Loerch2014,Marquardt2006,Elste2009,Krantz2016,Metelmann2014,BenArosh2021,Lee2014}.
In metastable regimes, the geometry of the underlying attractors determines the structure of rare switching.
For fixed-point attractors, escape is organized around localized regions of phase space.
By contrast, limit cycles and their basin boundaries are extended, so escape depends on the phase at which fluctuations approach the boundary.

Activated escape is traditionally described by Kramers' problem, where fluctuations induce exponentially rare transitions over a barrier through a saddle point~\cite{Kramers1940,Langer1969,Haenggi1990}.
This framework underlies reaction rates and noise-driven switching across physics~\cite{Devoret1985,Grabert1987,Horsthemke2006}, chemistry~\cite{Eyring1935,Grote1980,Pollak2005}, and biology~\cite{Hasty2000,Assaf2017}.
While Kramers-type ideas extend to nonequilibrium settings, in driven systems lacking detailed balance the activation cost is generally not reducible to a static potential barrier. 
Instead, escape is described by Freidlin-Wentzell large-deviation theory, in which transition probabilities are controlled by the action cost of fluctuation paths~\cite{Graham1984,Freidlin2004,Dykman2001,Luchinsky1999,Kamenev2011}.
Minimum-action methods, including the geometric minimum-action method, identify the dominant weak-noise escape path~\cite{Weinan2002,Weinan2004,Heymann2008,Weinan2010}.
At finite fluctuation strength, however, switching observables can receive measurable contributions from an ensemble of fluctuation trajectories, including subdominant escape corridors that are suppressed in the asymptotic weak-noise limit.

For extended attractors, the finite-fluctuation ensemble structure becomes especially important since the effective activation cost is typically phase dependent.
Classical large-deviation theory predicts phase-localized escape from periodic attractors, with a single minimum-action sector dominating in the weak-noise limit~\cite{Maier1993,Maier1996,Dykman1990,Dykman1994,Silchenko2003}.
In driven quantum systems, most studies of activated switching have instead focused on bistable fixed-point attractors~\cite{Andersen2020,Mavrogordatos2025,Mylnikov2025,Lee2025,Kamal2009} or on symmetry-protected settings reducible to an effective potential~\cite{Marthaler2006,Lin2015,Peano2010,Boness2025,Thompson2022,Thompson2025,Carde2026}.
It therefore remains unclear how finite quantum fluctuations populate competing escape channels between extended metastable attractors in a nonequilibrium setting, and whether their relative contributions can be reconstructed directly from trajectory data.

Here we address this question in a driven optomechanical resonator whose semiclassical dynamics support two coexisting stable limit cycles~\cite{Loerch2014,Marquardt2006,Elste2009,Krantz2016,Metelmann2014,BenArosh2021,Lee2014,Gao2015,Roque2020,Christou2021}.
Using quantum-jump trajectories across a controlled quantum-to-classical crossover, we identify metastable residence intervals with a hidden Markov model and reconstruct dominant and subdominant escape corridors directly from observed switching events.
We find a pronounced directional asymmetry: escape from the small-amplitude cycle proceeds through a single radial corridor and exhibits near-Arrhenius scaling, whereas escape from the large-amplitude cycle involves competing phase-localized corridors with distinct effective activation scales.
Together, the curvature in the fluctuation-dependent switching-rate scaling and the event-conditioned phase distributions identify finite-fluctuation multichannel quantum escape between extended attractors.

	
\section{Model and methods}
\label{sec:model_methods}

In this section we specify the microscopic model, derive the semiclassical limit, introduce the quantum-to-classical scaling used to tune fluctuation strength, and describe the quantum-jump trajectory framework used to extract switching events.
\subsection{Driven-dissipative model}
\label{subsec:driven_dissipative_model}
\begin{figure}[t]
    \centering
    \includegraphics[width=\columnwidth]{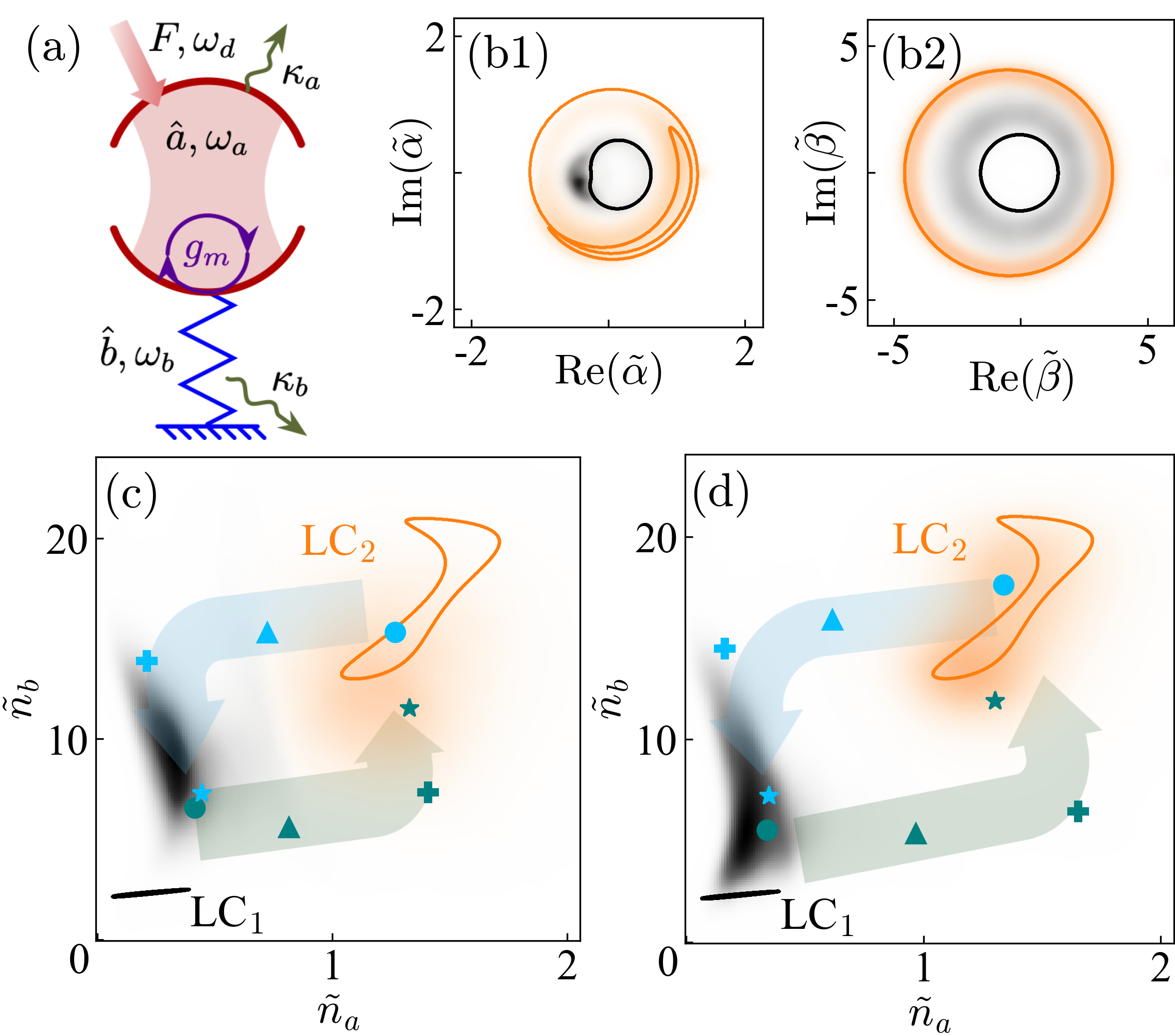}
    \caption{
        \textit{Coexisting limit cycles and quantum switching in an optomechanical resonator.}
        (a) Schematic of the driven optomechanical setup: a laser-driven cavity mode couples via radiation pressure to a mechanical oscillator.
        (b1,b2) The two coexisting limit cycles projected onto the complex coherent-amplitude phase spaces spanned by $(\mathrm{Re}(\tilde\alpha),\mathrm{Im}(\tilde\alpha))$ for the optical mode and $(\mathrm{Re}(\tilde\beta),\mathrm{Im}(\tilde\beta))$ for the mechanical mode.
        \LC{1} and \LC{2} obtained from the mean-field equations are represented as black and orange solid lines, respectively.
        Colored density maps show empirical stationary phase-space densities reconstructed from quantum-jump trajectory ensembles for $\aleph=9$, as described in Sec.~\ref{subsec:quantum_jump_trajectories}.
        Black and orange regions of the density map indicate the state assignment to \LC{1} and \LC{2}, respectively, as obtained from the hidden Markov model analysis.
        (c,d) Same quantities projected onto the rescaled population plane $(\tilde n_a,\tilde n_b)$ for (c) $\aleph=3$ and (d) $\aleph=9$.
        Green and blue transparent arrows indicate the predominant switching pathways extracted from event-conditioned trajectory ensembles.
        Green and blue symbols mark the phases at which snapshots of event-conditioned phase-space distribution are shown in Fig.~\ref{fig: event_aligned_wigner}, for \LC{1}$\to$\LC{2} and \LC{2}$\to$\LC{1} switching events, respectively.
    }
    \label{fig: setup_landscape}
\end{figure}

We consider a driven-dissipative optomechanical resonator consisting of a single optical mode $\hat a$ coupled via radiation pressure to a mechanical mode $\hat b$. 
	In a frame rotating at the drive frequency $\omega_d$, the Hamiltonian reads
	\begin{equation}
		\frac{\hat{\mathcal H}}{\hbar}
		=
		\Delta_a \hat a^\dagger\hat a
		+ \omega_b \hat b^\dagger\hat b
		+ g\,\hat a^\dagger\hat a(\hat b+\hat b^\dagger)
		+ F(\hat a+\hat a^\dagger)\;,
		\label{eq:Hamiltonian_PRL}
	\end{equation}
	where $\Delta_a=\omega_a-\omega_d$ is the optical detuning, $g$ is the single-photon optomechanical coupling, and $F$ is the coherent drive amplitude~\cite{Aspelmeyer2014}.
	Dissipation is described by a Lindblad master equation,
	\begin{equation}
		\frac{d\hat{\rho}}{dt}
		=
		\frac{1}{i\hbar}\bigl[\hat{\mathcal H}+\Lambda_b,\hat{\rho}\bigr]
		+ \kappa_a\,\mathcal D[\hat a]\hat{\rho}
		+ \kappa_b\,\mathcal D[\hat b]\hat{\rho}\;,
		\label{eq: master_eq}
	\end{equation}
	with $\mathcal D[\hat o]\hat\rho=\hat o\hat\rho\hat o^\dagger-\tfrac12\{\hat o^\dagger\hat o,\hat\rho\}$~\cite{Breuer2007}.
	The term $\Lambda_b=i(\kappa_b/4)(\hat b^{\dagger2}-\hat b^2)$ implements position damping for the mechanical quadratures within the Lindblad description~\cite{Dekker1977,Duffus2017,Wagner2026}.
	A schematic of the setup is shown in Fig.~\ref{fig: setup_landscape}~(a).

\subsection{Semiclassical dynamics}
\label{subsec:semiclassical_dynamics}
At the semiclassical level, the coherent amplitudes $\alpha=\langle\hat a\rangle$ and $\beta=\langle\hat b\rangle$ satisfy
\begin{equation}
    \begin{aligned}
        i\frac{d\alpha}{dt} &= (\Delta_a - i\kappa_a/2)\alpha + g\alpha(\beta+\beta^*) + F\;,\\
        i\frac{d\beta}{dt} &=  \omega_b \beta - i\kappa_b (\beta-\beta^*)/2 + g|\alpha|^2\;,
    \end{aligned}
    \label{eq:meanfield_GPE}
\end{equation}
which define a non-gradient nonlinear flow in a four-dimensional phase space spanned by $\mathrm{Re}(\alpha),\mathrm{Im}(\alpha),\mathrm{Re}(\beta),\mathrm{Im}(\beta)$.
We denote the corresponding semiclassical mode populations as $n_{a} \equiv \vert\alpha\vert^2$ and $n_{b} \equiv \vert\beta\vert^2$.
\subsection{Quantum-to-classical scaling}
\label{subsec:quantum_to_classical_scaling}
To systematically study the effect of quantum fluctuations, we introduce a one-parameter quantum-to-classical scaling manifold
\begin{equation}
	\begin{aligned}
		\alpha  &= \sqrt{\aleph}\,\tilde\alpha\;,& \qquad
		\beta   &= \sqrt{\aleph}\,\tilde\beta\;,& \\
		F       &= \sqrt{\aleph}\,\tilde F\;,& \qquad
		g       &= \tilde g/\sqrt{\aleph}\;.&
	\end{aligned}
	\label{eq:rescaling-aleph}
\end{equation}
Along this scaling manifold, the deterministic mean-field equations in Eq.~\eqref{eq:meanfield_GPE} remain invariant, while the dimensionless
parameter $\aleph$ tunes the quantum-to-classical crossover~\footnote{
Equivalently, for the rescaled operators
$\tilde{\hat a}=\hat a/\sqrt{\aleph}$ and
$\tilde{\hat b}=\hat b/\sqrt{\aleph}$, one has
$[\tilde{\hat a},\tilde{\hat a}^{\dagger}]
=[\tilde{\hat b},\tilde{\hat b}^{\dagger}]=1/\aleph$.
Thus, $\aleph^{-1}$ plays the role of an effective Planck constant.
}.
In the classical limit $\aleph \to \infty$, the coherent amplitudes scale
as $\alpha,\beta \sim \sqrt{\aleph}$, implying populations
$n_a,n_b \propto \aleph$.
Consequently, relative fluctuations are suppressed as $1/\sqrt{\aleph}$.
Thus, $\aleph$ directly controls the fluctuation strength and defines a
controlled semiclassical limit without shifting the classical
attractors~\cite{Seibold2020,Seibold2022,Nowoczyn2026}.
An equivalent scaling prescription that avoids tuning the microscopic 
coupling $g$, and is therefore more experimentally accessible, is given in App.~\ref{app:dual_scaling}.

Throughout this work, we fix the original-scale system parameters to
$\Delta_a=-0.7$, $\omega_b=1.0$, $\tilde g=0.35$, $\tilde F=0.2$, $\kappa_a=0.1$, and $\kappa_b=0.01$.
For these parameters, the deterministic flow admits two coexisting stable limit cycles. 
A phase diagram showing regions of coexistence in the surrounding parameter space is provided in App.~\ref{app:gpe_phase_diagram}.

\subsection{Quantum-jump trajectories}
\label{subsec:quantum_jump_trajectories}

Quantum switching dynamics are resolved using the quantum-jump, or Monte Carlo wave-function, unraveling of the Lindblad master equation Eq.~\eqref{eq: master_eq}~\cite{Dum1992,Dalibard1992,Moelmer1993,Carmichael1993}.
In this representation, the open-system dynamics are described by an ensemble of stochastic wave-function trajectories
$|\psi_r(t)\rangle$,
where $r$ labels the trajectory realization.
For a sufficiently large ensemble, the density matrix is recovered as
\begin{equation}
	\hat\rho(t)
	=
	\frac{1}{N_{\mathrm{traj}}}
	\sum_{r=1}^{N_{\mathrm{traj}}}
	|\psi_r(t)\rangle\langle\psi_r(t)| \;.
\end{equation}
Between quantum jumps, each trajectory evolves under the non-Hermitian effective Hamiltonian
\begin{equation}
	\hat H_{\mathrm{eff}}
	=
	\hat{\mathcal H}
	+
	\Lambda_b
	-
	\frac{i\hbar}{2}
	\left(
	\kappa_a \hat a^\dagger\hat a
	+
	\kappa_b \hat b^\dagger\hat b
	\right) \;,
\end{equation}
while stochastic photon and phonon emission events are generated by the jump operators $\hat a$ and $\hat b$, respectively.
The jump probability within a time interval $\Delta t$ is determined by the expectation value of the corresponding jump operator evaluated along the trajectory state $|\psi_r(t)\rangle$.

Throughout this work, observables are evaluated along individual quantum-jump trajectories.
We define the trajectory-resolved optical and mechanical populations as
\begin{equation}
\begin{aligned}
	n_a^{(r)}(t)
	&=
	\langle\psi_r(t)|
	\hat a^\dagger\hat a
	|\psi_r(t)\rangle\;,
	\\
	n_b^{(r)}(t)
	&=
	\langle\psi_r(t)|
	\hat b^\dagger\hat b
	|\psi_r(t)\rangle\;.
\label{eq:populations_single_traj}
\end{aligned}
\end{equation}
Analogously, the trajectory-resolved coherent amplitudes are defined as
\begin{equation}
\begin{aligned}
	\alpha^{(r)}(t)
	&=
	\langle\psi_r(t)|\hat a|\psi_r(t)\rangle\;,
	\\
	\beta^{(r)}(t)
	&=
	\langle\psi_r(t)|\hat b|\psi_r(t)\rangle\;.
\label{eq:coh_amplitudes_single_traj}
\end{aligned}
\end{equation}
%
The trajectory ensemble provides the central object of our analysis: it contains the metastable residence times, the rare switching events, and the phase-resolved geometry of the escape process.
We use it on three complementary levels.
(i)~Individual quantum-jump trajectories are represented by the trajectory-resolved populations and coherent amplitudes defined in Eqs.~\eqref{eq:populations_single_traj} and \eqref{eq:coh_amplitudes_single_traj}.
(ii)~Stationary population and phase-space densities are reconstructed by sampling these trajectory-resolved expectation values on a uniform time grid and accumulating them over the trajectory ensemble.
When constructed from coherent amplitudes, these densities are shown in the rescaled optical and mechanical phase spaces spanned by $\mathrm{Re}(\tilde\alpha),\mathrm{Im}(\tilde\alpha)$ and $\mathrm{Re}(\tilde\beta),\mathrm{Im}(\tilde\beta)$.
When constructed from populations, they are shown in the rescaled population plane $(\tilde n_a,\tilde n_b)$.
These distributions are empirical densities of trajectory-resolved observables, not Wigner quasiprobability distributions.
(iii)~Event-conditioned phase-space distributions and event-conditioned distributions of the optical phase $\phi_a$ are reconstructed from trajectory segments aligned relative to switching events identified by the hidden-Markov-model analysis.
Further technical details of the numerical implementation and reconstruction procedures are given in App.~\ref{app:qj}.

The quantum-to-classical scaling is implemented by scaling the system parameters $F$ and $g$ according to Eq.~\eqref{eq:rescaling-aleph}.
Throughout this work, we present trajectory expectation values in rescaled form, e.g. $\tilde n_j^{(r)} \equiv n_j^{(r)}/\aleph$, $j\in \{a,b\}$,
which makes results at different fluctuation strengths directly comparable and consistent with the scaling of the complex amplitudes in Eq.~\eqref{eq:rescaling-aleph}.

\section{Classical and quantum attractor landscape} 
\label{sec:landscape}

We characterize the deterministic semiclassical attractor structure of the driven-dissipative optomechanical resonator introduced in Sec.~\ref{sec:model_methods} and how this structure manifests in the finite-$\aleph$ quantum dynamics.

For the chosen parameters, the semiclassical equations of motion Eq.~\eqref{eq:meanfield_GPE} host two coexisting stable limit cycles, separated in phase space by an unstable periodic orbit that forms the classical basin boundary.
In the classical limit, $\aleph \to \infty$, quantum fluctuations vanish and the quantum dynamics reduces to the deterministic semiclassical motion, so that trajectories remain confined to the basin selected by their initial condition.
At finite $\aleph$, the semiclassical limit cycles are reflected in the quantum dynamics as metastable, long-lived oscillatory states.
Quantum fluctuations broaden each metastable state around its underlying limit cycle, producing transverse fluctuations about the orbit and phase diffusion along it~\cite{Nowoczyn2026,Seibold2020}.
On a much longer timescale, the same fluctuations induce rare transitions across the basin boundary, giving rise to interstate switching.
As a consequence of this metastability, the stationary state acquires support near both oscillatory states.
The scaling parameter $\aleph$ controls the strength of these fluctuation effects: increasing $\aleph$ suppresses local broadening, phase diffusion, and interstate switching, thereby recovering deterministic classical dynamics.
In the remainder of this paper, we refer to both the semiclassical limit cycles and their corresponding metastable long-lived oscillatory states in the quantum regime as \LC{1} and \LC{2}.

The two coexisting limit-cycle solutions of the mean-field equations, Eq.~\eqref{eq:meanfield_GPE}, are shown in Fig.~\ref{fig: setup_landscape}(b1-d) as black and orange lines, respectively. They are displayed in the optical and mechanical complex phase spaces (b1,b2) and in the rescaled $(\tilde n_a,\tilde n_b)$ population plane (c,d).
The two solutions form distinct closed loops, where  \LC{1} corresponds to a small-amplitude oscillation, while \LC{2} is a larger-amplitude oscillation.
\LC{2} exhibits a pronounced modulation of the instantaneous populations along the orbit, consistent with stronger amplitude-phase mixing.
In the complex optical plane spanned by $\Re(\tilde\alpha)$ and $\Im(\tilde\alpha)$, \LC{1} appears approximately circular and centered close to the origin, whereas \LC{2} traces a distorted circular orbit.
In the complex mechanical plane spanned by $\Re(\tilde\beta)$ and $\Im(\tilde\beta)$, both limit cycles appear nearly circular, but the center of \LC{2} is displaced from the origin.
The displacement and distortion of \LC{2} imply that different phases along the orbit sample different local linearized dynamics.
This produces phase-dependent amplitude-phase mixing, so fluctuations are locally amplified or suppressed depending on position along the cycle.
This angle-dependent stability is a key geometric feature underlying the phase-localized escape corridors analyzed below.

Stationary population and phase-space densities reconstructed from quantum-jump trajectories are shown in Fig.~\ref{fig: setup_landscape}~(b1-d), projected onto the optical and mechanical complex phase space in (b1,b2) and onto the $(\tilde n_a, \tilde n_b)$ population plane in (c,d), for small $\aleph=3$ in (c) and for larger $\aleph=9$ in (b1,b2,d).
The stationary phase-space densities for $\aleph=3$ are presented in App.~\ref{app:Wigner_stationary_state_and_quantum_melted_regime}.
For large $\aleph$, quantum fluctuations weakly perturb the classical dynamics, yielding thin annuli with probability density circulating along each cycle and a depleted region between them.
As $\aleph$ decreases, both annuli broaden and deform in an angle-dependent manner: the \LC{1} distribution expands outward, while the \LC{2} distribution contracts inward toward the basin boundary~\cite{Sarkar2024}.
For \LC{1}, the anisotropic landscape produces pronounced stationary-density hotspots associated with specific phase sectors, even though the deterministic attractor remains a single periodic orbit, see black line in Fig.~\ref{fig: setup_landscape}~(b1).
These density accumulations reflect phase-dependent fluctuation amplitudes and corresponding variations in local residence times along the cycle.
This angular structure foreshadows the phase-resolved escape corridors resolved in Sec.~\ref{sec:geom_escape}.

For sufficiently small $\aleph$, the strength of quantum fluctuations becomes comparable to (or larger than) the effective activation cost associated with transitions between \LC{1} and \LC{2}.
The broadened annuli then overlap and fill the region separating \LC{1} and \LC{2}, see App.~\ref{app:Wigner_stationary_state_and_quantum_melted_regime}.
In this strongly fluctuating regime, trajectories no longer relax toward well-separated regions of phase space, and the notion of distinct metastable oscillatory states loses its meaning.
We refer to this loss of the well-separated metastable attractor structure as a quantum-melted regime~\footnote{Here, \textit{quantum melting} denotes the fluctuation-induced loss of a well-separated metastable attractor structure. This usage is related to, but distinct from, Refs.~\cite{Nowoczyn2026,Seibold2020}, where the term refers to the degradation of long-lived periodic or quasiperiodic modes associated with a single attractor.
}.
In the following, we restrict to $\aleph \geq 2$, where metastability and separation of timescales are preserved.

\begin{figure}[t]
\centering
\includegraphics[width=\columnwidth]{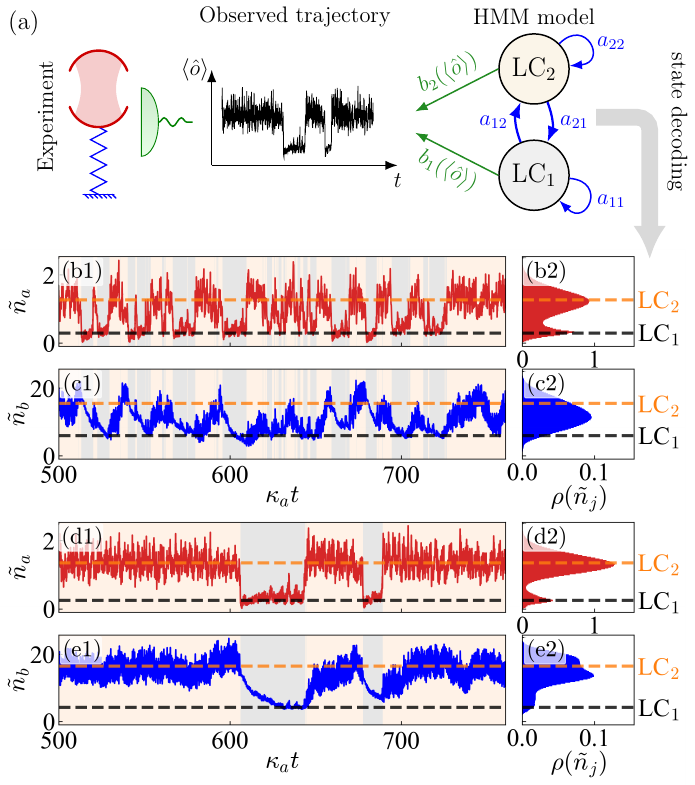}
\caption{
    \textit{State sequence inference, single-trajectory switching dynamics, and timescale separation of optical and mechanical populations.}
    (a) Trajectories are analyzed with a hidden Markov model (HMM) to extract the underlying state sequence.
    (b1, c1) Time traces of the trajectory-resolved rescaled expectation values $\tilde n_a^{(r)}$, $\tilde n_b^{(r)}$, obtained from a single quantum-jump trajectory at $\aleph=3$. Red and blue indicate the cavity and mechanical populations, respectively.
    (d1, e1) Same quantities for $\aleph=9$.
    Dashed horizontal lines indicate the mean populations of the metastable quantum states \LC{1} (black) and \LC{2} (orange).
    Black and orange shaded backgrounds show the state assignment to \LC{1} and \LC{2}, respectively, as obtained from the HMM analysis.
    Right panels display histograms for (b2, c2) $\aleph=3$, and (d2, e2) $\aleph=9$, accumulated over all trajectories.
    For stronger quantum fluctuations ($\aleph=3$), frequent switching and slow mechanical relaxation leads to population of transient states between \LC{1} and \LC{2}, washing out bimodality in $\tilde n_b$.
    For weaker fluctuations ($\aleph=9$), switching events are rare, allowing full relaxation onto each cycle and producing a clear bimodal distribution for both modes.
}
\label{fig: trajectory_switching}
\end{figure}
	
	
\section{Metastable switching dynamics and escape rates}
\label{sec:traj_states}
	
To resolve metastable switching between \LC{1} and \LC{2}, we analyze the stochastic dynamics generated by the quantum-jump unraveling of Eq.~\eqref{eq: master_eq}.
Individual quantum trajectories reveal three dynamical stages: exploration within a metastable neighborhood, rare fluctuation-induced escape, and post-switching relaxation toward the target cycle.
From these trajectories we extract residence times, survival functions, and directional switching rates.
Resolving these quantities as functions of $\aleph$ provides a rate-level quantification of how quantum fluctuations control metastable lifetimes and interstate switching.

To identify visits to each metastable state, we segment long trajectories into \LC{1} and \LC{2} intervals using a hidden Markov model (HMM)~\cite{Rabiner1986,Rabiner1989}, as detailed in App.~\ref{app:hmm}.
The HMM infers the most probable two-state sequence from the trajectory-resolved observables, assigning each time step to one of the two metastable states, see Fig.~\ref{fig: trajectory_switching}.
This coarse-grained two-state description is valid when intra-attractor relaxation is fast compared with inter-attractor switching, and it does not resolve the phase dynamics within each cycle.

	Fig.~\ref{fig: trajectory_switching}~(b1-e1) shows representative individual quantum-jump trajectories $\tilde{n}_a^{(r)}(t)$ and $\tilde{n}_b^{(r)}(t)$ for $\aleph=3$ and $\aleph=9$.
	For larger $\aleph$, trajectories exhibit long intervals of nearly periodic motion close to the semiclassical orbits of \LC{1} or \LC{2}, punctuated by rare stochastic transitions, see Fig.~\ref{fig: trajectory_switching}~(d1, e1).
	In the corresponding probability-density histograms of the rescaled populations, shown in Fig.~\ref{fig: trajectory_switching}~(d2, e2), this produces two well-separated peaks associated with \LC{1} and \LC{2}.
	The sub-peak structure in the histogram of $\tilde{n}_b$ arises from intrinsic oscillations of the population along each limit cycle and is not related to switching dynamics.
	For smaller $\aleph$, switching becomes more frequent and the system spends more time in the intermediate region between the two attractor basins, see Fig.~\ref{fig: trajectory_switching}~(b1-b2, c1-c2).
	
Independent of the specific value of $\aleph$, the mechanical population $\tilde{n}_b$ spends more time in this intermediate regime than the optical population $\tilde{n}_a$, see Fig.~\ref{fig: trajectory_switching}~(b2–e2).
This asymmetry becomes clear when examining individual switching events:
A transition between the two states is initiated by a rapid change in the optical-mode population, while the mechanical-mode population evolves more slowly and exhibits a transient \q{ring-down} before settling into the basin of the alternate attractor.
The optical mode thus governs the switching event, while the mechanical mode follows through slower relaxation dynamics.
This separation of timescales is directly visible in Fig.~\ref{fig: trajectory_switching} and Fig.~\ref{fig: setup_landscape}~(c,d), where rapid optical displacements are followed by slower mechanical relaxation.
This observation motivates reducing the full four-dimensional phase space to a slow manifold parametrized by the optical phase $\phi_a$, see Sec.~\ref{subsec:exit-point-clustering}.

To analyze survival statistics and directional switching rates, we use the state sequences inferred by the HMM to divide each trajectory into visits to the two metastable states.
For each visit to \LC{i}, we record the dwell time $T_{\text{dwell}}$ and estimate the corresponding survival function $S_i(t)$, defined as the probability that a trajectory remains in the basin of \LC{i} longer than $t$ after entering it, 
\begin{equation}
    S_i(t) = P(T_{\text{dwell}}>t \mid \text{entry into state}\;\LC{i})\;.
    \label{eq:survival_functin_main}
\end{equation}
Representative results of the empirical survival functions $S^{\text{emp}}_i(t)$, obtained using the Kaplan-Meier estimator~\cite{Kaplan1958}, are shown in App.~\ref{app:survival_analysis}.
For both \LC{1} and \LC{2}, we focus on the long-time behavior of $S^{\text{emp}}_i(t)$.
This excludes short stochastic \q{flickering} events that occur when trajectories linger near the basin boundary.
It also excludes escape during the mechanical \q{ring-down} phase discussed above, which produces additional curvature in $S^{\text{emp}}_i(t)$ at intermediate times before convergence to an exponential tail, see App.~\ref{app:survival_analysis}.
At sufficiently long times, the empirical survival functions exhibit an exponential tail,
\begin{equation}
    S^{\text{emp}}_i(t)\sim e^{-k_{ij}t}\;,
    \label{eq:emp_survival_exponential_form}
\end{equation}
consistent with Markovian switching dynamics and activated escape governed by a single asymptotic rate $k_{ij}$ for each direction \LC{i} to \LC{j} at fixed $\aleph$.
To extract the directional switching rates $k_{12}$ and $k_{21}$, we perform a maximum-likelihood fit to the dwell-time data, based on a conditional survival model
\begin{equation}
    S_i(t \vert T_{\text{dwell}}\geq t_0) = \frac{S_i(t)}{S_i(t_0)} = e^{-k_{ij}(t-t_0)} \;,
\end{equation}
where $t_0$ is chosen beyond the transient regime, such that the model only describes survival at long times, where the system has relaxed into the basin and $S_i^{\text{emp}}(t)$ exhibits a clear single-exponential behavior.
The fitted models together with the empirical survival functions are shown in App.~\ref{app:survival_analysis}.
	
\begin{figure}[t]
    \centering
    \includegraphics[width=\columnwidth]{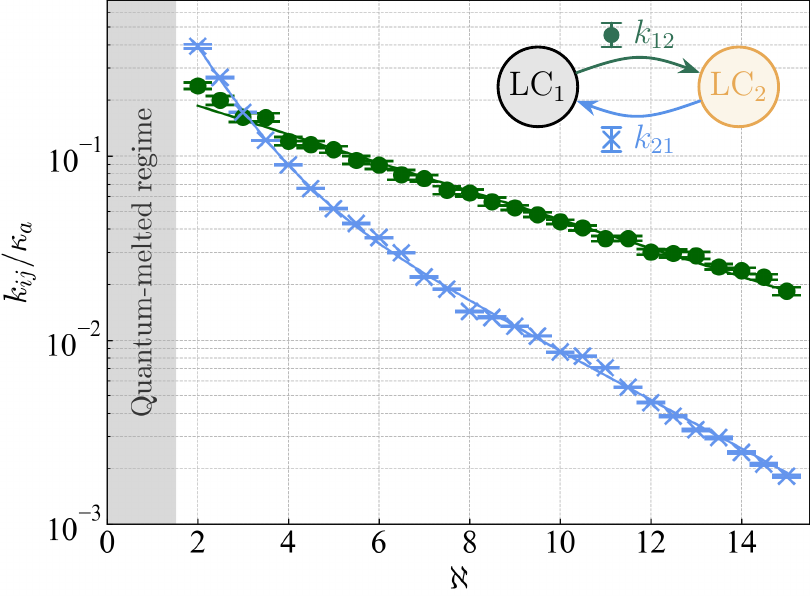}
    \caption{
        \textit{Multichannel scaling of switching rates.}
        Directional switching rates $k_{ij}/\kappa_a$ between \LC{i} and \LC{j} as functions of the scaling parameter $\aleph$.
        The \LC{1}$\to$\LC{2} switching rate $k_{12}$, shown in green, follows an approximately single-exponential Arrhenius dependence over the explored range of $\aleph$, consistent with activation through a single dominant escape corridor.
        The green solid line shows the fit to a single-exponential form for $k_{12}(\aleph)$ as given in Eq.~\eqref{eq: single_exponential_rate}, with fitted exponent $S_{12} = 0.178 \pm 0.004$ and prefactor $A_{12}=0.267\pm 0.010$.
        By contrast, the \LC{2}$\to$\LC{1} switching rate $k_{21}$, shown in blue, exhibits pronounced curvature when plotted as $\log k_{21}$ versus $\aleph$, following a biexponential form that reflects the coexistence of geometrically distinct switching pathways contributing at finite fluctuation strength.
        The blue solid line shows the fit to a biexponential form for $k_{21}(\aleph)$ as given in Eq.~\eqref{eq: biexponential_rate}, with fitted exponents $S_{21}^{\mathrm{ph}} = 0.303 \pm 0.008$ and $S_{21}^{\mathrm{amp}} = 1.057 \pm 0.035$, and prefactors $A_{21}^{\mathrm{ph}}=0.180\pm0.015$ and $A_{21}^{\mathrm{amp}}=2.454\pm0.140$, illustrating the crossover from phase-averaged to minimal-action escape as $\aleph$ increases.
        The gray-shaded area marks the quantum-melted regime at $\aleph \leq 1.5$, where quantum fluctuations effectively destroy the metastability of the coexisting states and switching rates cannot be extracted, see App.~\ref{app:Wigner_stationary_state_and_quantum_melted_regime}.
    }
    \label{fig: rates_vs_aleph}
\end{figure}

In Fig.~\ref{fig: rates_vs_aleph}, we show the directional switching rates as functions of the scaling parameter $\aleph$.
The rate $k_{12}$ follows single-exponential scaling over the explored range, indicating that escape from \LC{1} is controlled by one dominant fluctuation corridor.
In contrast, $k_{21}$ shows pronounced curvature in $\log k_{21}$ versus $\aleph$, revealing that escape from \LC{2} is not governed by a single effective activation scale at finite fluctuations.
Instead, the rate decomposes into competing contributions from multiple geometric escape corridors whose relative weights vary with $\aleph$.
This finite-fluctuation multichannel structure is consistent with escape from extended periodic attractors and does not signal a breakdown of activation or large-deviation theory~\cite{Maier1993,Maier1996}.
In Sec.~\ref{sec:geom_escape}, we resolve these phase-localized corridors directly from event-conditioned quantum trajectories.
In Sec.~\ref{sec:action_based_interpretation}, we then connect this trajectory-resolved geometry to an effective action-based description of the observed rate scaling.

The observed timescales of the switching dynamics are also reflected in the Liouvillian spectrum.
In the metastable regime, the slowest nonzero Liouvillian mode corresponds to relaxation between the two long-lived oscillatory states.
Projecting the dynamics onto the HMM-inferred two-state manifold gives the effective rate matrix
\begin{equation}
\mathbf K =
\begin{pmatrix}
-k_{12} & k_{21}\\
 k_{12} & -k_{21}
\end{pmatrix}\;,
\end{equation}
whose nonzero relaxation rate is
\begin{equation}
    \Lambda_{\mathrm{eff}} = k_{12}+k_{21}\;.
\end{equation}
Thus, the trajectory-extracted switching rates determine the slow Liouvillian relaxation scale.
Additional details on the metastable reduction and its relation to the Liouvillian spectrum are provided in App.~\ref{app:liouvillian_gap}.

\section{Phase-space geometry of multichannel escape}
\label{sec:geom_escape}

\begin{figure*}[t]
    \centering
    \includegraphics[width=2\columnwidth]{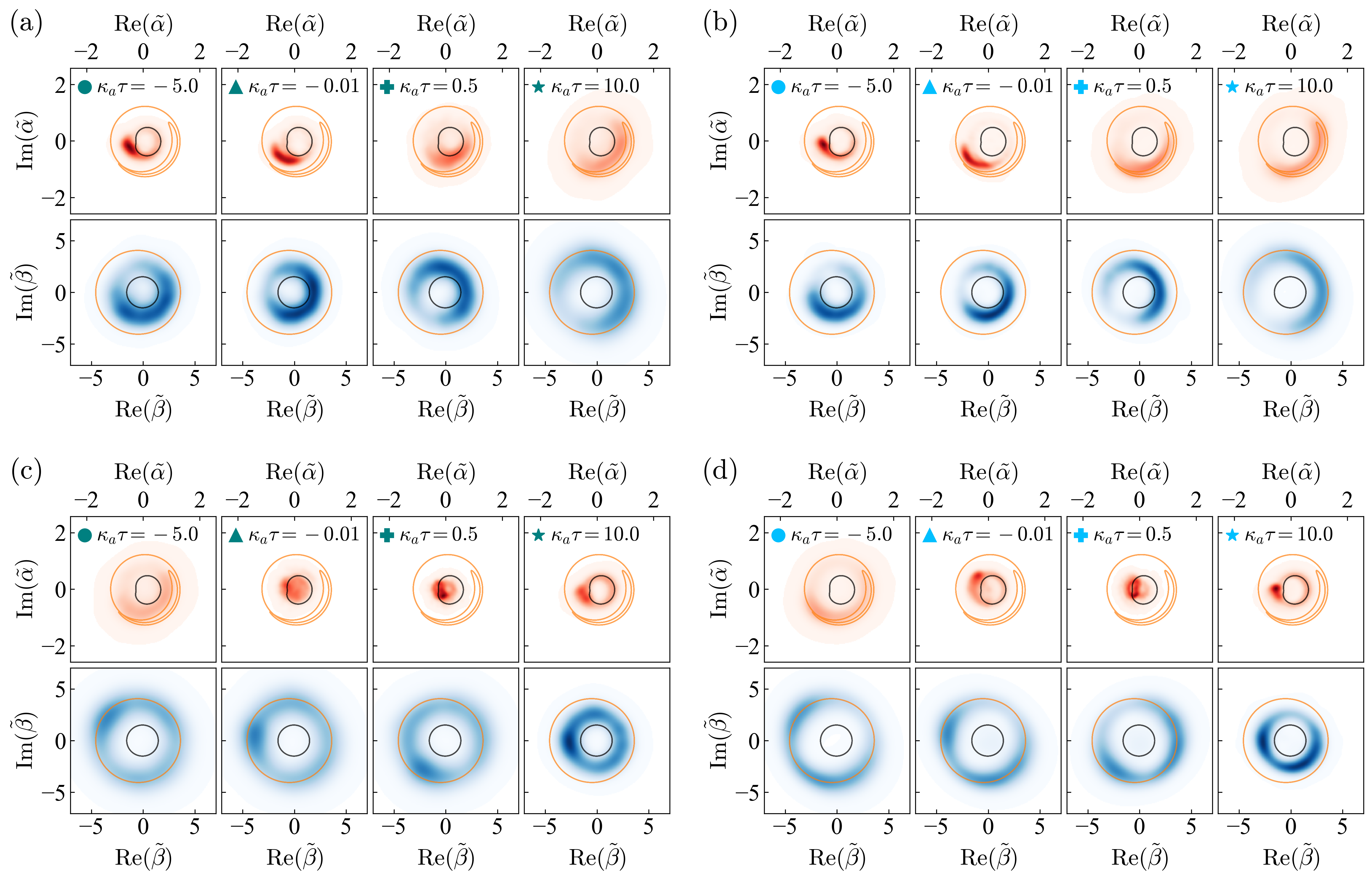}
    \caption{
        \textit{Phase-space distributions of event-conditioned trajectories during limit-cycle switching events.}
        Phase-space distributions conditioned on \LC{1}$\to$\LC{2} switching events for (a) $\aleph=3$ and (b) $\aleph=9$, and conditioned on \LC{2}$\to$\LC{1} switching events for (c) $\aleph=3$ and (d) $\aleph=9$.
        In each panel, upper and lower rows show the optical and mechanical phase spaces, respectively, with black and orange lines indicating the classical limit cycles \LC{1} and \LC{2}, respectively.
        Columns correspond to increasing times relative to the switching event, illustrating the system before the switch ($\kappa_a\tau=-5.0$), approaching the basin boundary ($\kappa_a\tau=-0.01$), shortly after crossing ($\kappa_a\tau=0.5$), and after relaxation and ring-down onto the other limit cycle ($\kappa_a\tau=10.0$).
        The snapshots at switching time, $\kappa_a\tau=-0.01$, reveal a qualitative difference in the fluctuation-dependence of the phase-distribution during the switch for \LC{1}$\to$\LC{2} and \LC{2}$\to$\LC{1} directions.
        At the switching time, $\kappa_a\tau=-0.01$, \LC{1}$\to$\LC{2} events remain phase-localized for both fluctuation strengths, whereas \LC{2}$\to$\LC{1} events become broadly distributed in phase as fluctuations increase.
    }
    \label{fig: event_aligned_wigner}
\end{figure*}

The rate analysis of Sec.~\ref{sec:traj_states} revealed a pronounced directional asymmetry: escape from \LC{1} is consistent with a single dominant corridor, whereas escape from \LC{2} shows a finite-fluctuation multichannel signature.
We now resolve the corresponding escape geometry directly from event-conditioned quantum trajectories, identifying the fluctuation corridors through which escape occurs and visualizing how they organize along the extended basin boundary.
This event-resolved perspective shows how the attractor structure controls multichannel escape, while the weak-noise limit remains governed by a single dominant pathway.
	
\subsection{Event-resolved escape geometry}
\label{subsec: event resolved escape geometry}
To extract typical escape geometries, we isolate segments from many individual quantum-jump trajectories around switching events and align them such that the switch occurs at $\tau\equiv t-t_{\mathrm{switch}}=0$.
Here $t_{\mathrm{switch}}$ is defined by the crossing of the basin boundary separating \LC{1} and \LC{2}.
We include only trajectory segments with a minimum dwell time $t_{\mathrm{dwell}}=15.0/\kappa_a$ in the initial basin before the switch and $t_{\mathrm{dwell}}=10.0/\kappa_a$ in the target basin after the switch.
This removes short flickering events near the separatrix, ensures relaxation near the initial cycle before escape, and resolves the post-switch mechanical ring-down.
For each \LC{i}$\to$\LC{j} transition, we record the rescaled populations $(\tilde n_a^{(r)},\tilde n_b^{(r)})$ and complex amplitudes $(\tilde\alpha^{(r)},\tilde\beta^{(r)})$ and accumulate the corresponding conditional ensemble.
The resulting event-conditioned phase-space distributions provide an empirical reconstruction of the escape geometry: dominant fluctuation corridors appear as concentrated probability tubes in phase space.

In Fig.~\ref{fig: event_aligned_wigner}, snapshots of the event-conditioned phase-space distributions are shown for the switch \LC{1}$\to$\LC{2}, for (a)~$\aleph=3$, (b)~$\aleph=9$, and for the switch \LC{2}$\to$\LC{1}, for (c)~$\aleph=3$, (d)~$\aleph=9$.
The displayed snapshot times correspond to four characteristic stages of the transition.
Before the switch, $\kappa_a \tau = -5.0$, both optical and mechanical modes are confined within the basin of the initial limit cycle.
At the switching time, $\kappa_a\tau=-0.01$, the optical-mode population shifts toward the basin of the target limit cycle and the system crosses the unstable periodic boundary.
Shortly after the switch, $\kappa_a\tau=0.5$, the optical mode has reached the population level of the target limit cycle, while the mechanical mode exhibits only minor changes.
This separation of timescales reflects the effective reduction of the switching dynamics to the optical coordinate discussed in Sec.~\ref{sec:traj_states}.
After ring-down, $\kappa_a\tau=10.0$, the mechanical population has relaxed onto the target limit cycle.
These stages are also highlighted in Fig.~\ref{fig: setup_landscape}, with symbols corresponding to the snapshot times shown in Fig.~\ref{fig: event_aligned_wigner}.
Complementary movies of the full time evolution are provided in the Supplemental Material.

The event-conditioned distributions exhibit a directional asymmetry consistent with the observed asymmetry in switching-rate scaling.
At $\kappa_a\tau=-0.01$, \LC{1}$\to$\LC{2} escape is sharply phase-localized for both $\aleph=3$ and $\aleph=9$, see Fig.~\ref{fig: event_aligned_wigner}~(a,b).
This fluctuation-independent phase selectivity indicates that a single dominant corridor controls this switching direction over the explored range.
For \LC{2}$\to$\LC{1}, escape is phase-selective at weaker fluctuations ($\aleph=9$), but becomes broadly distributed over a wide angular sector at stronger fluctuations ($\aleph=3$), see Fig.~\ref{fig: event_aligned_wigner}~(c,d).
This broadening shows that additional escape sectors acquire statistical weight at finite fluctuation strength.
The geometric broadening of the event-conditioned distributions provides the direct phase-space counterpart of the curvature observed in $k_{21}(\aleph)$ and identifies the multichannel origin of the rate scaling.
	
	
\subsection{Exit-point clustering and fluctuation corridors}
\label{subsec:exit-point-clustering}
	
A complementary perspective on the escape geometry in phase space is provided by the event-conditioned distributions of the optical phase $\phi_a=\arg(\alpha)$.
As discussed in Sec.~\ref{sec:traj_states}, the switching dynamics in the full four-dimensional phase space can be effectively reduced to a slow manifold parametrized by this phase, enabling a phase-resolved analysis along the cycles.
As in Sec.~\ref{subsec: event resolved escape geometry}, we construct the event-conditioned distributions by isolating trajectory segments around switching events and aligning them relative to the switching time $t_{\text{switch}}$.
For both directions \LC{i}$\to$\LC{j}, we collect all phases $\phi_a^{(r)}$ in the conditioned ensemble to reconstruct event-conditioned phase histograms $P(\phi_a, \tau \mid \LC{i}\to\LC{j})$, normalized to unity over $\phi_a$ for each $\tau$.
These distributions thus reflect the conditional probability that the system is at phase $\phi_a$ at time $\tau$ relative to a switching event \LC{i} $\to$ \LC{j} at $\tau=0$.
We only include trajectory segments with a minimum dwell time $t_{\text{dwell}}=15.0/\kappa_a$ in the initial basin before the switch and $t_{\text{dwell}}=2.0/\kappa_a$ in the target basin after the switch.
This removes short-timescale flickering near the separatrix, ensures sufficient pre-switch dynamics to resolve the escape process, and provides a short post-switch window to confirm that the system switched basins.
		
	\begin{figure}[t]
		\centering
		\includegraphics[width=1\linewidth]{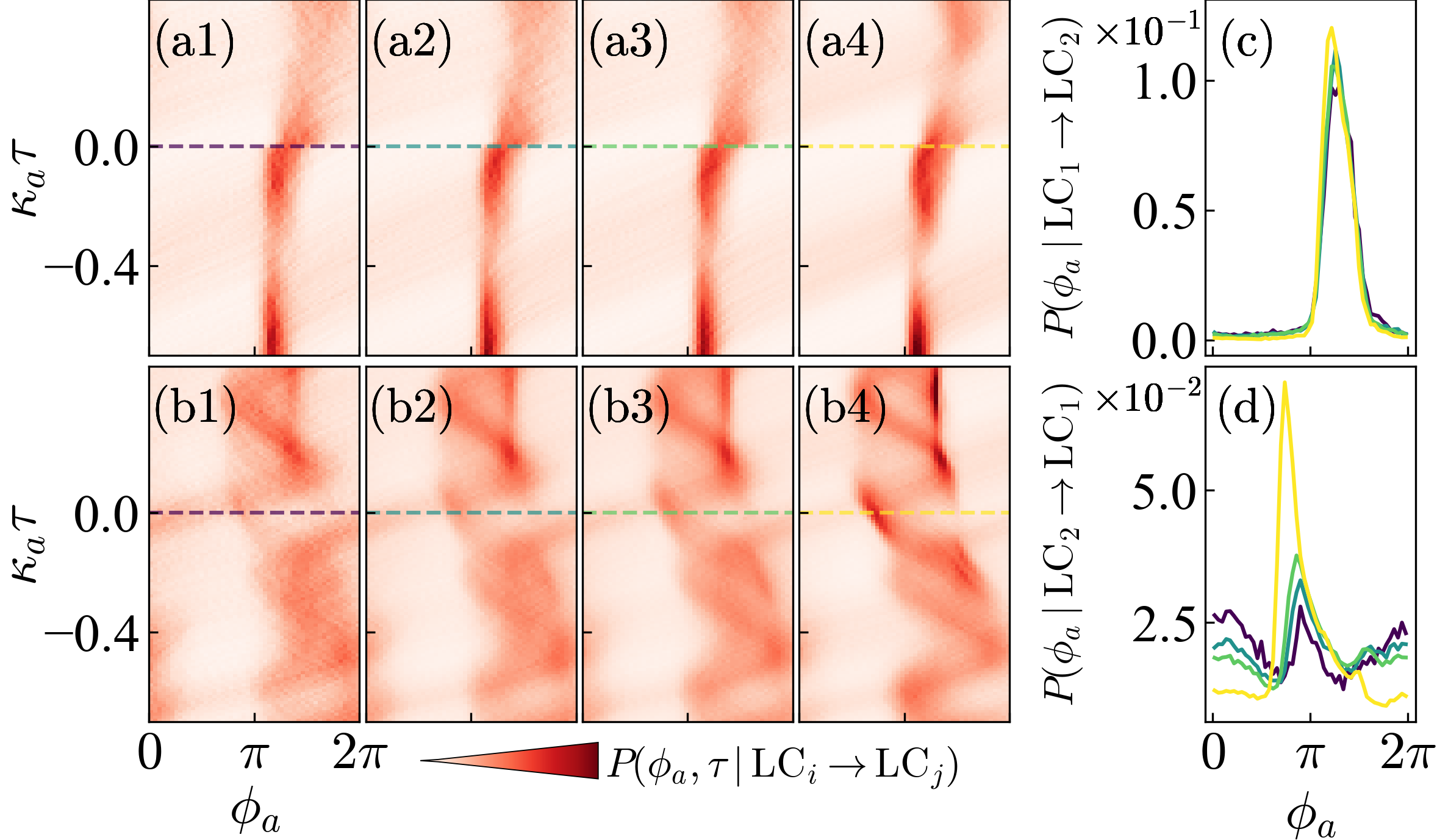}
		\caption{
			\textit{Event-conditioned phase distributions during switching.}
			Normalized event-conditioned phase distributions $P(\phi_a,\tau \mid \LC{i}\to\LC{j})$ for \LC{1}$\to$\LC{2} in the upper row and \LC{2}$\to$\LC{1} in the lower row.
			Columns~$1-4$ correspond to increasing fluctuations strengths, $\aleph=2.5$, $\aleph=3.0$, $\aleph=3.5$, and $\aleph=9.0$.
			(a1-a4) For \LC{1}$\to$\LC{2}, the escape-phase distributions show strong phase localization and well-defined gateway phases, independent of the fluctuation strength.
			(b1-b4) For \LC{2}$\to$\LC{1}, the escape-phase distribution displays strong phase localization at small fluctuations (b4), and an emerging additional broad region of escape phases as fluctuations increase (b4$\to$b1).
			(c, d) Cuts at $\kappa_a\tau=0$, corresponding to the colored dashed lines in (a1-a4) and (b1-b4), displaying the instantaneous escape-phase distributions at switching time $P(\phi_a,\tau=0 \mid \LC{i}\to\LC{j})$, for (c) \LC{1}$\to$\LC{2} and (d) \LC{2}$\to$\LC{1}.
		}
		\label{fig: phase_hazard}
	\end{figure}
	
	In Fig.~\ref{fig: phase_hazard}, we show $P(\phi_a, \tau \mid \LC{1}\to\LC{2})$ in (a1-a4) and $P(\phi_a, \tau \mid \LC{2}\to\LC{1})$ in (b1-b4), with columns corresponding to increasing scaling parameter $\aleph$.
	The colored dashed lines highlight the time $\tau=0$, at which the system crosses the basin boundary.
	In Fig.~\ref{fig: phase_hazard}~(c, d) we show the corresponding instantaneous escape-phase distribution $P(\phi_a \mid \LC{i}\to\LC{j}) \equiv P(\phi_a, \tau=0 \mid \LC{i}\to\LC{j})$ for the same values of $\aleph$.
	
	For switching events \LC{1}$\to$\LC{2}, the event-conditioned phase distribution reveals that escape from \LC{1} is strongly localized within a narrow angular sector, see Fig.~\ref{fig: phase_hazard}~(a1-a4, c).
	This sharp phase clustering indicates a single dominant escape corridor governed by radial fluctuations.
	Importantly, the angular width of the escape sector remains essentially unchanged across the explored values of $\aleph$, showing that the same escape corridor controls switching as the fluctuation strength varies.
	This is consistent with the single-exponential Arrhenius scaling observed for the switching rate $k_{12}$.
	%
	
	By contrast, escape from \LC{2} exhibits a broader and multimodal angular structure, see Fig.~\ref{fig: phase_hazard}~(b1-b4, d).
	In particular, the shape of the event-conditioned phase distribution depends sensitively on the fluctuation strength.
	For large $\aleph$, the distribution displays a single sharp peak, corresponding to a strongly phase-gated radial crossing of the basin boundary.
	As $\aleph$ decreases, an additional broad region of phases contributes to the distribution.
	In this regime, phase-dependent amplitude-phase mixing becomes increasingly relevant, allowing escape through angular sectors where the effective radial activation cost differs along the orbit.
	The coexistence of these angular escape sectors directly mirrors the observed biexponential dependence of $k_{21}$ on fluctuation strength and the resulting curvature in $\log(k_{21})$, and provides a structural explanation for the multichannel scaling.

	Since trajectories explore the basins of \LC{1} and \LC{2} nonuniformly, the event-conditioned phase distributions shown in Fig.~\ref{fig: phase_hazard} are biased by the stationary phase distribution within the initial basin and therefore do not directly reflect the phase-dependent escape propensity or the action costs associated with different escape paths.
	Formally, the event-conditioned phase distributions conditioned on the transition \LC{i} $\to$ \LC{j} can be decomposed as
	\begin{equation}
		\begin{aligned}
			P(\phi_a,\tau \mid \LC{i}\!\to\!\LC{j})
			&=
			P_{\mathrm{stat}}(\phi_a \mid \mathrm{LC}(\tau))  \\
			&\qquad \times
			P(\tau \mid \phi_a, \LC{i}\!\to\!\LC{j}) \;,
		\end{aligned}
	\end{equation}
	where $P_{\text{stat}}(\phi_a \mid \LC{i})$ and $P_{\text{stat}}(\phi_a \mid \LC{j})$ denote the stationary phase distributions within the basins of \LC{i} and \LC{j}, respectively, and
	\begin{align}
		\text{LC}(\tau) =
		\begin{cases}
			\LC{i} & \text{for } \tau \leq 0 \\
			\LC{j} & \text{for } \tau > 0
		\end{cases} \;.
	\end{align}
	In \appref{app:hazard}, we show the normalized event-conditioned phase distributions $P(\tau\mid\phi_a,\LC{i}\to\LC{j})$ and the phase-conditioned escape hazard
	\begin{equation}
		h(\phi_a)\equiv P(\tau=0 \mid \phi_a, \LC{i}\to\LC{j}) \;,
	\end{equation}
	both obtained by normalizing the event-conditioned distributions with the stationary phase distributions.

	The clustering of escape phases represents the quantum analogue of geometric focusing in classical escape from periodic attractors~\cite{Maier1993,Maier1997}.
	Here, multichannel escape refers to the coexistence of geometrically distinct fluctuation pathways originating from the same metastable basin, whose relative contributions vary with fluctuation strength.
	The curvature observed in the scaling of the switching rates $k_{ij}$ with $\aleph$ reflects the competition between phase-localized escape channels with distinct activation costs, whose relative weights vary with fluctuation levels.
	

\section{Action-based interpretation}
\label{sec:action_based_interpretation}

Having resolved the escape corridors directly from event-conditioned trajectories, we now connect this finding to the observed rate scaling through an effective activation-action picture.
The exponents introduced below are empirical activation scales extracted from the $\aleph$-dependence of the directional switching rates.
They should be understood as finite-fluctuation diagnostics of the phase-resolved escape landscape, rather than as actions obtained from an explicit Freidlin-Wentzell variational calculation.
This interpretation is consistent with large-deviation theory for escape from periodic attractors, where the activation cost is distributed along the orbit and the weak-fluctuation limit selects the minimum-action sector~\cite{Maier1996,Berglund2014,Cruz2018}.

The escape geometries identified in Sec.~\ref{sec:geom_escape} leave a direct quantitative imprint on the switching rates in Fig.~\ref{fig: rates_vs_aleph}.
For \LC{1}$\to$\LC{2}, the rate is well described over the explored range by a single effective exponential,
\begin{equation}
    k_{12}(\aleph)
    =
    A_{12} e^{-S_{12}\aleph}\;,
    \label{eq: single_exponential_rate}
\end{equation}
consistent with activation through one dominant escape corridor characterized by an effective action cost $S_{12}$.
By contrast, for \LC{2}$\to$\LC{1} escape involves multiple phase sectors, and the corresponding rate displays pronounced curvature in $\log k_{21}$ versus $\aleph$.
Over the explored range, this curvature is captured by
\begin{equation}
    k_{21}(\aleph)
    =
    A_{21}^{\rm ph} e^{-S_{21}^{\rm ph}\aleph}
    +
    A_{21}^{\rm amp} e^{-S_{21}^{\rm amp}\aleph} \;,
    \label{eq: biexponential_rate}
\end{equation}
with effective action costs $S_{21}^{\rm ph}$ and $S_{21}^{\rm amp}$, where $S_{21}^{\rm ph}<S_{21}^{\rm amp}$, and prefactors $A_{21}^{\rm ph}$ and $A_{21}^{\rm amp}$, where $A_{21}^{\rm ph} < A_{21}^{\rm amp}$.
The lower-action contribution corresponds to the phase-gated corridor that dominates in the weak-fluctuation, large-$\aleph$ limit.
The higher-action contribution becomes visible at stronger fluctuations because its larger prefactor compensates part of the exponential suppression over the accessible range of $\aleph$.
In this effective two-channel description, the prefactors encode non-exponential channel factors, including the phase-space measure of the corresponding escape sector and local fluctuation prefactors. 
Thus, the curvature of $k_{21}(\aleph)$ reflects a redistribution of probability among escape sectors, rather than a breakdown of activated switching or large-deviation scaling~\cite{Luchinsky1997,Khasin2009,Chan2007}.

This interpretation is supported by the phase-resolved diagnostics of Sec.~\ref{sec:geom_escape}.
For \LC{1}$\to$\LC{2}, the exit-phase distribution remains narrowly localized as $\aleph$ is varied, matching the single-effective-action form of Eq.~\eqref{eq: single_exponential_rate}.
For \LC{2}$\to$\LC{1}, the exit-phase distribution changes qualitatively with fluctuation strength: at large $\aleph$ it is concentrated near a dominant phase-gated sector, while at smaller $\aleph$ an additional broad angular sector acquires weight. 
This broadening signals a crossover toward amplitude-like escape, where transverse fluctuations enable exits over a wider range of phases.
The biexponential form in Eq.~\eqref{eq: biexponential_rate} is therefore a compact parametrization of the finite-fluctuation channel mixture observed directly in the trajectory ensemble.

More generally, escape from an extended attractor can be represented, at the level of large-deviation asymptotics, as a phase integral over local escape sectors~\footnote{For a general overview of optimal-path and large-deviation approaches to noise-induced transitions in nonequilibrium dynamical systems, see Ref.~\cite{Forgoston2018,Heller2024}.},
\begin{equation}
    k(\aleph)
    \sim
    \frac{1}{2\pi}
    \int_0^{2\pi} d\phi\,
    A(\phi)
    \exp[-\aleph S(\phi)] \;,
    \label{eq: phase_action}
\end{equation}
where $\phi$ parametrizes position along the limit cycle, $S(\phi)$ is a phase-resolved activation cost, and $A(\phi)$ encodes the statistical weight of escape trajectories near phase $\phi$~\cite{Dykman1994,Cruz2018,Meerson2019}.
In the weak-fluctuation limit $\aleph\to\infty$, the integral is dominated by
\begin{equation}
    S_{\min}=\min_\phi S(\phi)\;,
    \label{eq:min_action}
\end{equation}
yielding $k(\aleph)\sim e^{-\aleph S_{\min}}$.
At finite fluctuation strength, however, multiple phase sectors can contribute simultaneously.
The instantaneous slope of the rate defines an effective activation scale,
\begin{equation}
    S_{\mathrm{eff}}(\aleph)
    =
    -\frac{d}{d\aleph}\log k(\aleph)
    =
    \int d\phi\, w(\phi,\aleph)S(\phi)\;,
    \label{eq:effective_action}
\end{equation}
 where
\begin{equation}
    w(\phi,\aleph)
    =
    \frac{A(\phi)e^{-\aleph S(\phi)}}
    {\int d\phi\,A(\phi)e^{-\aleph S(\phi)}}
\end{equation}
is the normalized weight associated with the escape sector at phase $\phi$.
Thus, $S_{\mathrm{eff}}(\aleph)$ is a fluctuation-dependent weighted average over the phase-resolved activation landscape.
As $\aleph$ increases, the weight concentrates on the minimum-action sector; at smaller $\aleph$, broader or higher-action sectors can contribute appreciably.
This provides the action-based explanation for the multichannel scaling of $k_{21}$ while remaining consistent with the single-exponential survival-function tails at fixed $\aleph$, see Sec.~\ref{sec:traj_states} and App.~\ref{app:survival_analysis}.

\section{Conclusion and outlook}
\label{sec:conclusion and outlook}

We have shown that quantum fluctuations induce phase-resolved multichannel switching between coexisting limit cycles in driven-dissipative quantum systems, and that the associated escape geometry can be reconstructed directly from quantum-jump trajectories.
In the semiclassical limit, the dynamics is organized by stable limit-cycle attractors.
At finite fluctuation strength, the corresponding oscillatory quantum states become metastable and undergo rare transitions across an extended periodic basin boundary.
In contrast to escape from fixed points, switching between extended attractors is governed by the geometry of the phase-space flow and intrinsically phase dependent.

Using quantum-jump simulations across a controlled quantum-to-classical crossover in a driven optomechanical resonator, we analyzed the switching statistics and escape geometry between the two metastable long-lived oscillatory states corresponding to two coexisting limit cycles in the underlying semiclassical dynamics.
The trajectory ensemble provides direct access to residence times, directional switching rates, event-conditioned phase-space distributions, and exit-phase statistics.
This approach is complementary to weak-noise minimum-action methods: while Freidlin-Wentzell theory identifies the asymptotically dominant escape path, finite-fluctuation trajectory ensembles reveal how subdominant escape corridors acquire measurable statistical weight.
Within this framework, a hidden Markov model provides a minimal inference layer for extracting the slow two-state switching dynamics from noisy quantum trajectories.

We find a pronounced asymmetry between the two switching directions.
Escape from the small-amplitude cycle is dominated by a single radial corridor and displays single-effective-action scaling over the explored fluctuation range.
By contrast, escape from the large-amplitude cycle involves competing phase-localized escape sectors with distinct effective activation scales.
The curvature in the corresponding switching-rate scaling, together with the broadening and multimodality of the event-conditioned phase distributions, identifies the finite-fluctuation multichannel character of this escape process.
In the weak-fluctuation limit, the statistical weight concentrates onto a single dominant corridor, consistent with the minimum-action picture.

More generally, our results show that escape from nonequilibrium extended attractors can be intrinsically phase resolved and organized by attractor geometry.
Because the mechanism relies on generic features of driven-dissipative systems with neutral phase directions and periodic basin boundaries, it is not restricted to optomechanical implementations.
The trajectory-based reconstruction developed here can be applied to other quantum nonlinear oscillators, including superconducting, electromechanical, and photonic platforms in which individual stochastic trajectories are experimentally accessible.

These findings suggest that stochastic switching in nonequilibrium quantum systems can be engineered through attractor geometry, dissipation, and fluctuation structure, opening a route toward controlling rare transitions in complex driven-dissipative dynamics.

\begin{acknowledgments}
	We thank D. K. J. Bone{\ss} for constructive feedback on the manuscript. CN and LM acknowledge funding by the Cluster of Excellence \q{Advanced Imaging of Matter} (EXC 2056), Project No. 390715994. The project is co-financed by ERDF of the European Union and by \q{Fonds of the Hamburg Ministry of Science, Research, Equalities and Districts (BWFGB)}. KS acknowledges funding from the Deutsche Forschungsgemeinschaft (DFG) via project number 449653034.
\end{acknowledgments}
	
\section*{Appendices}

\appendix

In this Appendix, we present the methods and supplementary results underlying our analysis of multichannel escape between coexisting quantum limit cycles and its interpretation in terms of phase-resolved activation.
It is organized as follows.
In \appref{app:gpe_phase_diagram}, we analyze the semiclassical dynamics and map out the phase diagram.
\appref{app:qj} describes the quantum-jump implementation used to simulate the system dynamics.
In \appref{app:dual_scaling}, we detail the quantum-to-classical scaling introduced in the main text and present an alternative scaling prescription which does not require tuning of the optomechanical coupling constant.
\appref{app:Wigner_stationary_state_and_quantum_melted_regime} provides supplementary plots of the quantum stationary phase-space densities, for scaling parameters used in the main text and for the quantum-melted regime.
In \appref{app:hmm} we detail the Hidden Markov model (HMM) used for segmentation of the trajectories, and discuss its application and limitations.
\appref{app:survival_analysis} explains how the directional switching rates $k_{ij}$ are extracted from the survival statistics of the two metastable oscillatory states.
In \appref{app:hazard}, we provide a complementary analysis to the results of Sec.~\ref{sec:geom_escape}, normalizing the event-conditioned phase distributions by the stationary phase distribution.
In \appref{app:liouvillian_gap}, we detail how metastability is reflected in the Liouvillian spectrum and explain the connection between the directional switching rates $k_{ij}$ and the Liouvillian gap.

\section{Mean-field phase diagram}
\label{app:gpe_phase_diagram}

In this Appendix we characterize the deterministic phase diagram associated with the mean-field equations Eq.~\eqref{eq:meanfield_GPE} of the optomechanical system, in order to place the working point used throughout the main text in a broader dynamical landscape.
Optomechanical systems exhibit a variety of dynamical behaviors, including multistability across different parameter regimes~\cite{Gao2015,Roque2020,Christou2021}.
Our goal is not an exhaustive classification, but to demonstrate that the coexistence of two stable limit cycles persists over an extended parameter region, and that the working point lies well within this region, far from any bifurcation.
The analysis also provides insight into the surrounding dynamical regimes.
\paragraph{Parameter scan and numerical protocol.}
Our classification of the dynamical phases is based on the long-time asymptotic behavior of the mean-field equations, see Eq.~\eqref{eq:meanfield_GPE}.
These nonlinear equations can admit multiple stable attractors, including fixed points and limit cycles. When multiple attractors coexist, phase space decomposes into their respective basins of attraction, such that the long-time behavior is selected by the initial condition.

To construct the phase diagram, we vary the rescaled drive strength $\tilde F$ and the optical detuning $\Delta_a$ while holding all other mean-field parameters fixed to the values used in the main text (including $\omega_b$, $\tilde J$, and $\kappa_{a,b}$).
For each pair $(\Delta_a,\tilde F)$ we integrate the mean-field equations from a set of $N_{\text{init}}=157$ distinct initial conditions, where the initial cavity field amplitude $\tilde\alpha(0)=\tilde\alpha_r(0)+i\tilde\alpha_i(0)$ uniformly samples phase space for $\abs{\tilde\alpha}\leq 5$, while the mechanical initial condition is fixed to $\tilde\beta(0)=0$.
Time evolution is performed using an explicit Runge-Kutta solver (RK45) to final time $t_f=3\times 10^3/\kappa_a$.
Dynamical observables are extracted from the late-time window $t\in[t_f-30/\kappa_a, t_f]$ to eliminate transients.
\paragraph{Identification and classification of coexisting attractors.}
For each trajectory we compute peak-to-peak amplitudes in the late-time window,
\begin{equation}
	\begin{aligned}
		\Delta \tilde\alpha_r &\equiv \max_t \mathrm{Re}\,\tilde\alpha(t)
		- \min_t \mathrm{Re}\,\tilde\alpha(t) \;, \\
		\Delta \tilde\beta_r  &\equiv \max_t \mathrm{Re}\,\tilde\beta(t)
		- \min_t \mathrm{Re}\,\tilde\beta(t)\;.
	\end{aligned}
\end{equation}
as simple order parameters that distinguish convergence to a fixed point from convergence to an extended attractor supporting time-dependent trajectories.
We classify a mode as \emph{static} if its peak-to-peak amplitude falls below the threshold $\epsilon_{\text{osc}}=5\times10^{-3}$, and as \emph{dynamical} otherwise.
To further discriminate between distinct solutions that may exhibit identical peak-to-peak amplitudes, e.g. multiple coexisting fixed points, we additionally compute the late-time mean occupations,
\begin{equation}
	\begin{aligned}
		\mathrm{mean}(\tilde n_a) &\equiv \mathrm{mean}(\abs*{\tilde\alpha(t)}^2)\;, \\
		\mathrm{mean}(\tilde n_b) &\equiv \mathrm{mean}(\abs*{\tilde\beta(t)}^2)\;.
	\end{aligned}
\end{equation}
To identify distinct attractors, we collect the set of distinct late-time outcomes obtained from the different initial conditions at fixed $(\Delta_a,\tilde F)$, grouping trajectories whose diagnostic vectors $(\Delta\tilde\alpha_r,\Delta\tilde\beta_r,\mathrm{mean}(\tilde n_a),\mathrm{mean}(\tilde n_b))$ agree within the tolerance $\epsilon_{\text{tol}}=10^{-2}$.
The number of distinct clusters determines the number of coexisting attractors, while the corresponding peak-to-peak amplitudes determine their type (fixed point versus limit cycle). This procedure yields a robust and reproducible phase-diagram classification.

In the parameter regime studied, the distinct dynamical attractors differ strongly in oscillation amplitude; therefore amplitude-based clustering unambiguously separates coexisting limit cycles.
%
In all cases identified as dynamical, the corresponding Fourier spectra exhibit a single dominant frequency with harmonics and no additional incommensurate peaks, confirming purely periodic behavior.
Within the explored parameter region, we find no evidence of other asymptotic dynamical behavior such as quasiperiodicity or chaotic dynamics.

\paragraph{Bistable limit-cycle region and working point.}
\begin{figure}
	\centering
	\includegraphics[width=0.99\columnwidth]{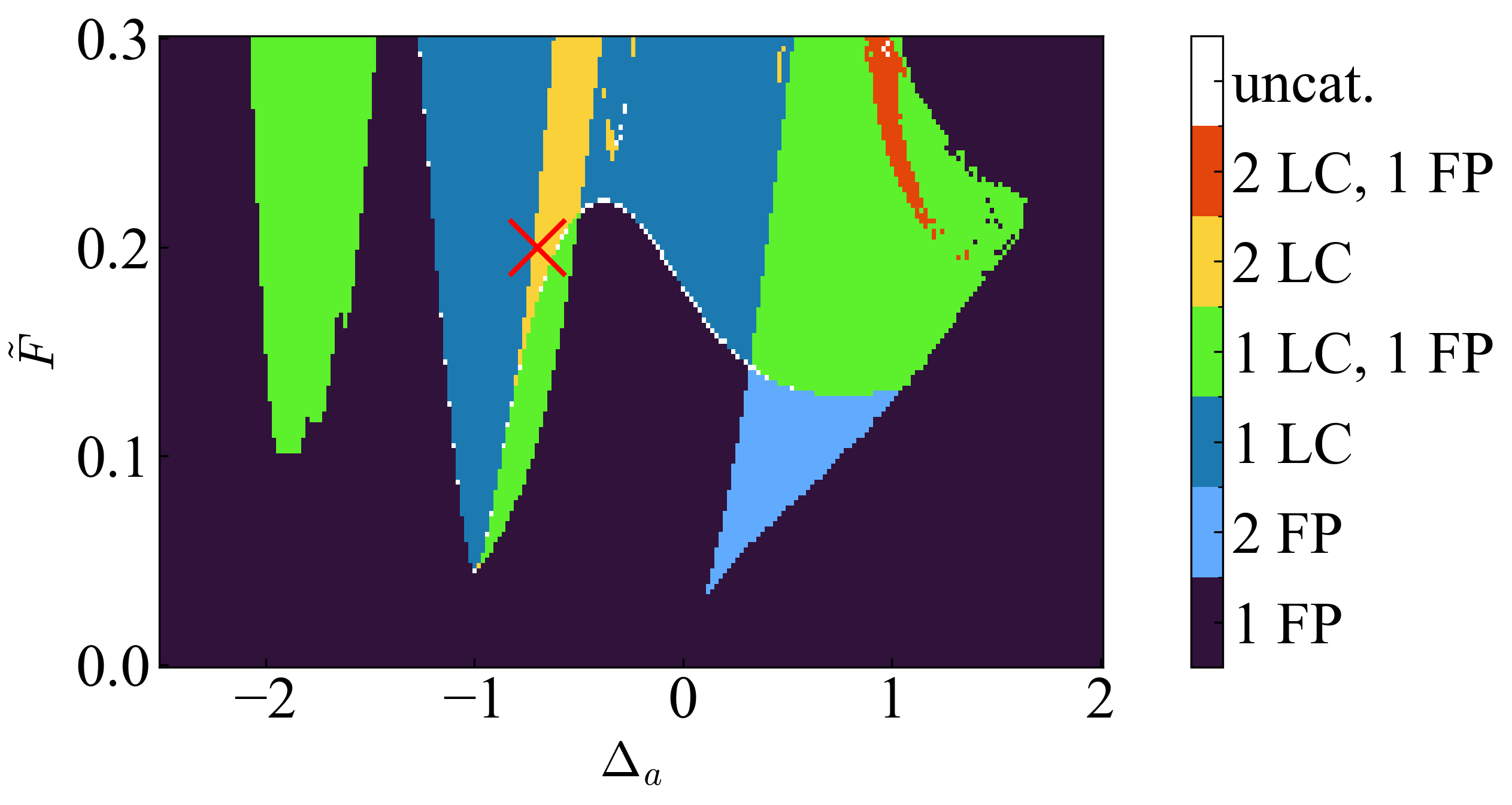}
	\caption{
		\textit{Deterministic phase diagram of the mean-field (GPE) dynamics in the $(\Delta_a,\tilde F)$ plane.}
		Each pixel is obtained by integrating the mean-field equations from a grid of initial conditions for $\tilde\alpha(0)$ and classifying the long-time behavior using late-time peak-to-peak amplitudes and mean populations $(\Delta\tilde\alpha_r,\Delta\tilde\beta_r,\mathrm{mean}(\tilde n_a),\mathrm{mean}(\tilde n_b))$.
		Colors indicate the number and type of distinct attractors observed across initial conditions:
		(i) a single fixed point (\q{1FP}),
		(ii) two coexisting fixed points (\q{2FP}),
		(iii) a single limit cycle (\q{1LC}),
		(iv) coexistence of one fixed point and one limit cycle (\q{1LC, 1FP}),
		(v) coexistence of two distinct limit cycles (\q{2LC}),
		(vi) coexistence of two distinct limit cycles and a single fixed point (\q{2LC, 1FP}),
		and (vii) uncategorized regimes (\q{uncat.}) with coexistence of more than $3$ attractors or critical slowing down at the phase boundaries.
		The red marker indicates the working point used in the main text ($\Delta_a=-0.7, \tilde F= 0.2$), which lies well within the region of two coexisting limit cycles and away from its boundaries.
	}
	\label{fig:app:gpe_phase_diagram}
\end{figure}
Figure~\ref{fig:app:gpe_phase_diagram} shows the resulting phase diagram in the $(\Delta_a,\tilde F)$ plane.
The yellow region labeled as \q{$\text{2 LC}$} corresponds to parameter sets for which two distinct limit cycles are found across the scanned initial conditions. The remaining regions correspond to a single limit cycle (\q{1LC}), a single fixed point (\q{1FP}), two coexisting fixed points (\q{2FP}), coexistence of a fixed point with a limit cycle (\q{1LC, 1FP}), or coexistence of a fixed point and two limit cycles (\q{2LC, 1FP}).
White points labeled \q{$\text{uncat.}$} correspond to parameter sets that are not classified into one of the displayed phases, either due to the coexistence of more than $3$ static and/or dynamical attractors, or due to critical slowing down at the phase boundaries, where the final time $t_f$ of our simulations is not large enough to reach the asymptotic behavior.

The working point used throughout the main text is ($\Delta_a=-0.7, \tilde F= 0.2$), highlighted by the red marker in Fig.~\ref{fig:app:gpe_phase_diagram}. It lies well inside the two limit-cycle region and away from its boundaries, ensuring that the switching mechanisms analyzed in the main text are not associated with proximity to a bifurcation.

\section{Quantum-jump implementation}
\label{app:qj}

Individual quantum trajectories are simulated by unraveling the Lindblad master equation Eq.~\eqref{eq: master_eq} into stochastic wave-function dynamics using the quantum-jump (Monte Carlo wave-function) method~\cite{Dalibard1992,Dum1992,Moelmer1993,Carmichael1993}.
In this representation, the open-system dynamics are described by an ensemble of stochastic wave-function trajectories
$|\psi_r(t)\rangle$,
where $r$ labels the trajectory realization.
For a sufficiently large ensemble, the density matrix is recovered as
\begin{equation}
	\hat\rho(t)
	=
	\frac{1}{N_{\mathrm{traj}}}
	\sum_{r=1}^{N_{\mathrm{traj}}}
	|\psi_r(t)\rangle\langle\psi_r(t)| \;.
\end{equation}
%
%
Between quantum jumps, each trajectory evolves under the non-Hermitian effective Hamiltonian
\begin{equation}
	\hat H_{\mathrm{eff}}
	=
	\hat{\mathcal H}
	+
	\Lambda_b
	-
	\frac{i\hbar}{2}
	\left(
	\kappa_a \hat a^\dagger\hat a
	+
	\kappa_b \hat b^\dagger\hat b
	\right) \;,
\end{equation}

where $\hat{\mathcal H}$ is defined in Eq.~\eqref{eq:Hamiltonian_PRL} of the main text. The term $\Lambda_b=i(\kappa_b/4)(\hat b^{\dagger2}-\hat b^2)$ implements position damping for the mechanical quadratures within the Lindblad description~\cite{Dekker1977,Duffus2017,Wagner2026}.
%

%
Quantum emissions are incorporated as stochastic jumps associated with the collapse operators $\hat a$ and $\hat b$. 
At each small time step $\Delta t$, there is a certain probability of the system undergoing a discontinuous quantum jump.
The channel-resolved jump probabilities are
\begin{align}
	p_a &= \kappa_a\,\langle\psi_r(t)|\hat a^\dagger\hat a|\psi_r(t)\rangle\,\Delta t,\\
	p_b &= \kappa_b\,\langle\psi_r(t)|\hat b^\dagger\hat b|\psi_r(t)\rangle\,\Delta t,
\end{align}
with total probability $p_{\mathrm{tot}}=p_a+p_b\ll 1$.
In the algorithm, a uniform random number $\mu\in[0,1)$ is drawn at each step.
If $\mu<p_{\mathrm{tot}}$, a jump occurs and the channel is selected according to the relative weights $p_a/p_{\mathrm{tot}}$ and $p_b/p_{\mathrm{tot}}$.
Upon a jump, the state is updated as
\begin{equation}
	|\psi_r(t+\Delta t)\rangle =
	\frac{\hat a\,|\psi_r(t)\rangle}{\|\hat a\,|\psi_r(t)\rangle\|}
	\;\;\text{or}\;\;
	|\psi_r(t+\Delta t)\rangle = \frac{\hat b\,|\psi_r(t)\rangle}{\|\hat b\,|\psi(t)\rangle\|}\;,
\end{equation}
after which the trajectory continues with deterministic evolution under $\hat{H}_{\text{eff}}$ until the next stochastic jump.
Since between jump events, the state obeys the non-unitary Schrödinger equation
\begin{equation}
	\frac{d}{dt}|\psi_r(t)\rangle = -\frac{i}{\hbar}\hat{H}_{\mathrm{eff}}|\psi_r(t)\rangle \;,
\end{equation}
its norm decays according to the no-emission probability. Between jumps, the state is therefore renormalized to maintain $\langle\psi_r|\psi_r\rangle=1$.
The time step $\Delta t$ is chosen adaptively such that $p_{\mathrm{tot}}\ll1$ throughout the evolution, ensuring the validity of the first-order jump probabilities while resolving the deterministic dynamics on timescales set by $\kappa_a^{-1}$, $\kappa_b^{-1}$, and the intrinsic oscillation periods of the limit cycles.

Trajectories are initialized as coherent states, with amplitudes randomly sampled within the accessible phase-space region and propagated for a transient time $t^\ast = 500/\kappa_a$. After discarding the transient dynamics, each trajectory effectively samples the full steady-state distribution.
Observables are evaluated along individual quantum-jump trajectories as described in the main text.
All observables and switching statistics are computed from the post-transient portion of the trajectories. Over long integration times, $t_f-t^*=2000/\kappa_a$, the stochastic dynamics explores the neighborhood of each attractor, providing near-uniform coverage along the corresponding limit cycles.
The Hilbert space of each mode is truncated at sufficiently large occupation numbers to ensure convergence, and switching statistics are obtained from ensembles of long trajectories whose total duration exceeds the characteristic switching times by several orders of magnitude.

\section{Theoretical and experimentally implementable semiclassical scaling}
\label{app:dual_scaling}

To probe the switching dynamics between the two metastable oscillatory states in a controlled semiclassical limit, we introduce a dimensionless scaling parameter $\aleph$ that increases the typical mode populations while preserving the deterministic mean-field phase portrait defined by Eq.~\eqref{eq:meanfield_GPE}.
This scaling controls the strength of quantum fluctuations without modifying the underlying metastable attractors.

We employ two complementary but dynamically equivalent scaling prescriptions: (A) a formal theoretical scaling used in the main text and simulations, and (B) an experimentally implementable adjoint scaling that keeps the microscopic device parameters fixed.

\smallskip
\noindent\textbf{(A) Theoretical $\aleph$ scaling used in the main text.}
\par
We introduce rescaled amplitudes
\begin{equation}
	\alpha=\sqrt{\aleph}\; \tilde\alpha\;,\qquad
	\beta=\sqrt{\aleph}\; \tilde\beta\;,
\end{equation}
together with
\begin{equation}
	F=\sqrt{\aleph}\;\tilde F\;,\qquad
	g=\tilde g/\sqrt{\aleph}\;,
	\label{eq:app:rescaling-aleph}
\end{equation}
while keeping the remaining microscopic parameters (e.g. $\Delta_a,\omega_b,\kappa_a,\kappa_b$) fixed. 
Substituting Eq.~\eqref{eq:app:rescaling-aleph} into the mean-field equations Eq.~\eqref{eq:meanfield_GPE} shows that all deterministic drift terms scale uniformly as $\sqrt{\aleph}$, such that dividing by $\sqrt{\aleph}$ yields equations of motion for $(\tilde\alpha,\tilde\beta)$ independent of $\aleph$.
In the full master-equation treatment, this scaling is implemented solely at the level of the system parameters, $F=\tilde F \sqrt{\aleph}$, $g=\tilde g/\sqrt{\aleph}$.
Under this transformation, the typical mode populations scale proportionally to $\aleph$, consistent with the mean-field amplitudes scaling as $\sqrt{\aleph}\tilde\alpha$ and $\sqrt{\aleph}\tilde\beta$.
By contrast, the quantum noise associated with the Lindblad dissipators remains set by the bare decay rates and does not acquire any additional $\aleph$ dependence. Consequently, absolute fluctuations remain of order unity, while relative fluctuations are suppressed as $1/\sqrt{\aleph}$.
The parameter $\aleph$ therefore controls the strength of quantum effects and defines a controlled semiclassical limit $\aleph\to\infty$, in which the dynamics approach the deterministic mean-field description~\cite{Seibold2020,Seibold2022,Nowoczyn2026}.
Since the scaling parameter controls the excitation number of the modes, in the large-$\aleph$ limit the commutator $[\hat a, \hat a^{\dagger}]=1$ becomes negligible compared to $\expval{\hat a^{\dagger}\hat a} \propto \aleph$. The scaling parameter can therefore be interpreted as defining an effective Planck constant, $\hbar_{\mathrm{eff}}\sim 1/\aleph$.

\smallskip
\noindent\textbf{(B) Experimentally implementable adjoint scaling at fixed $g$.}
\par
A direct implementation of Eq.~\eqref{eq:app:rescaling-aleph} would require tuning the microscopic coupling $g$, which is typically fixed by device fabrication and not adjustable in situ.
However, the choice of scaling prescription leading to a well-defined classical limit is not unique.
Here we introduce an alternative scaling that preserves the same $\aleph$-independent deterministic drift while keeping $g$ fixed.

As before, we use the amplitude rescaling
\begin{equation}
	\alpha=\sqrt{\aleph}\;\tilde\alpha\;,\qquad
	\beta=\sqrt{\aleph}\;\tilde\beta\;.
\end{equation}
The system parameters scale as
\begin{align}
	\Delta_a &= \sqrt{\aleph}\;\tilde\Delta_a\;, &
	\omega_b &= \sqrt{\aleph}\;\tilde\omega_b\;, \nonumber\\
	\kappa_a &= \sqrt{\aleph}\;\tilde\kappa_a\;, &
	\kappa_b &= \sqrt{\aleph}\;\tilde\kappa_b\;, \nonumber\\
	F &= \aleph\;\tilde F \;,
\end{align}
together with a rescaled time $t'=\sqrt{\aleph}\,t$.
In particular, the optomechanical coupling $g$ remains unchanged. Substituting these transformations into Eq.~\eqref{eq:meanfield_GPE} verifies that, when expressed in $(\tilde\alpha,\tilde\beta,t')$, the deterministic drift is again independent of $\aleph$.
%

\section{Stationary phase-space densities and quantum melting of metastability}
\label{app:Wigner_stationary_state_and_quantum_melted_regime}
\begin{figure}
	\centering
	\includegraphics[width=0.95\columnwidth]{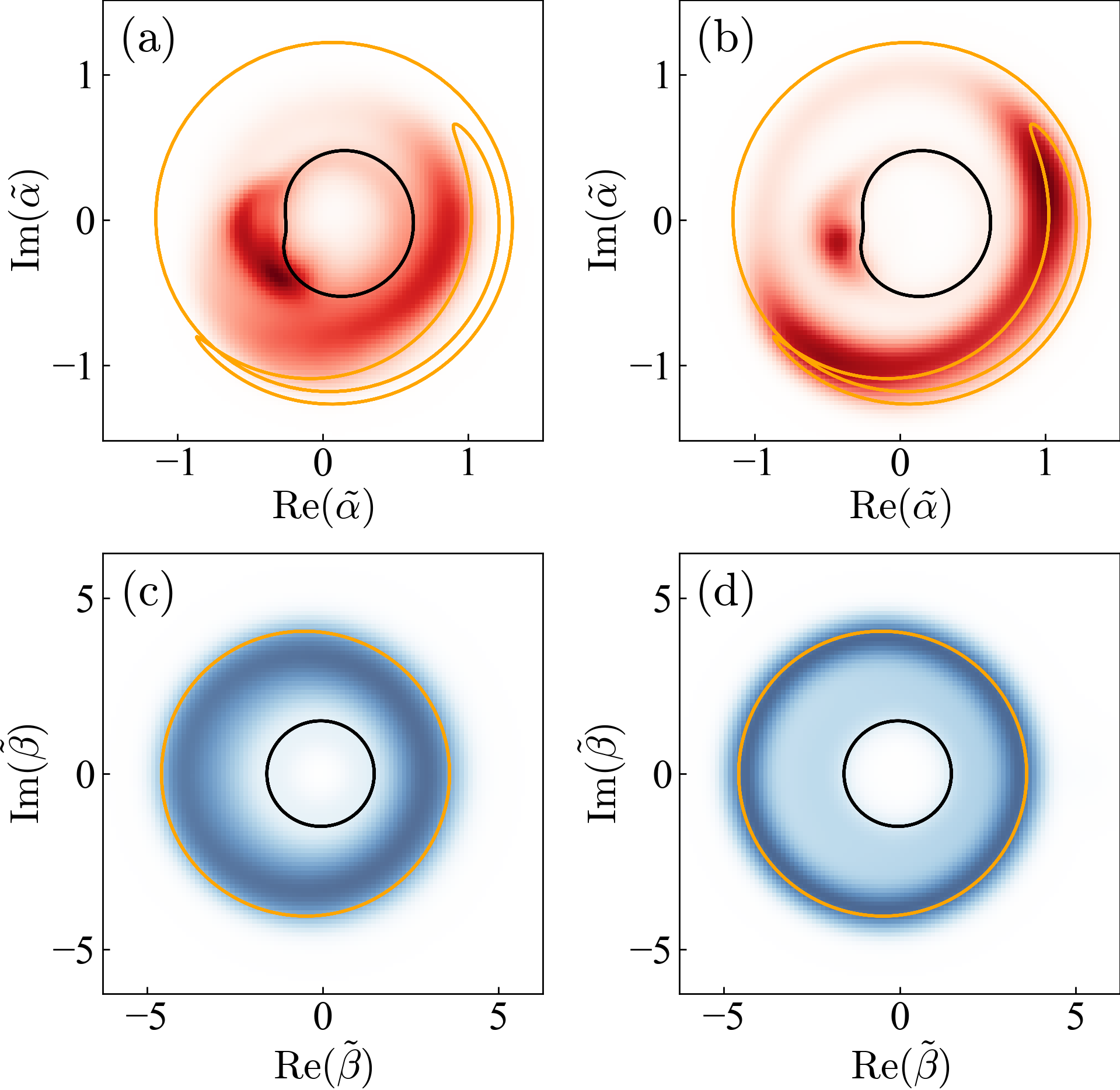}
	\caption{
		\textit{Stationary phase-space densities.}
		Stationary phase-space densities  reconstructed from quantum-jump trajectory ensembles for (a, c) $\aleph=3$ and (b, d) $\aleph=9$, projected onto the optical and the mechanical phase space in red and blue, respectively.
		Black and orange solid lines correspond to the two coexisting mean-field solutions \LC{1} and \LC{2}, respectively.
	}
	\label{fig:app:quantum_stationary_state_wigner_function_projections}
\end{figure}

\paragraph{Stationary phase-space densities}
As discussed in Sec.~\ref{sec:landscape}, quantum fluctuations enable switching between the two long-lived oscillatory states corresponding to the coexisting semiclassical limit cycles.
In Fig.~\ref{fig:app:quantum_stationary_state_wigner_function_projections}, we show the stationary phase-space densities reconstructed from quantum-jump trajectory ensembles as described in Sec.~\ref{subsec:quantum_jump_trajectories} for (a, c) $\aleph=3$ and (b, d) $\aleph=9$.
Panels (a) and (b) show projections onto the complex phase space of the cavity mode, for $\aleph=3$ and $\aleph=9$, respectively.
For both strong ($\aleph=3$) and weak ($\aleph=9$) fluctuations, the coexisting attractors remain clearly resolved.
Increasing quantum fluctuations broaden the annular distributions, and both the \LC{1} and \LC{2} distributions move closer to the basin boundary separating the attractors.
For $\aleph=3$, the directional switching rates between \LC{1} and \LC{2} are comparable, with $k_{12}/k_{21}\approx 1$. In contrast, for weak quantum fluctuations, $\aleph=9$, the rates differ by roughly an order of magnitude, with $k_{12}/k_{21}\approx 4$. As a consequence, in the weak-fluctuation regime the stationary-state distribution is dominated by \LC{2}.
Projections onto the complex phase space of the mechanical mode are shown in panels (c) and (d).
Here, the separation between the two basins is less pronounced; however, the underlying reason differs qualitatively between the two fluctuation regimes.
In the case of strong quantum fluctuations, see panel (c), the low occupation of \LC{1} arises from the ring-down dynamics of the mechanical mode discussed in Sec.~\ref{sec:traj_states}. While the optical-mode population changes rapidly when the system switches between the two attractors, the mechanical-mode population evolves on a slower timescale. As a result, the mechanical mode spends most of the time in the region between the two basin minima.
By contrast, for weak quantum fluctuations, see panel (d), the low occupation of \LC{1} is primarily a consequence of the asymmetry in the directional switching rates. As discussed above, switching from \LC{1} to \LC{2} is more likely than the reverse process, leading to a stationary distribution dominated by \LC{2}.

\begin{figure}
	\centering
	\includegraphics[width=0.95\columnwidth]{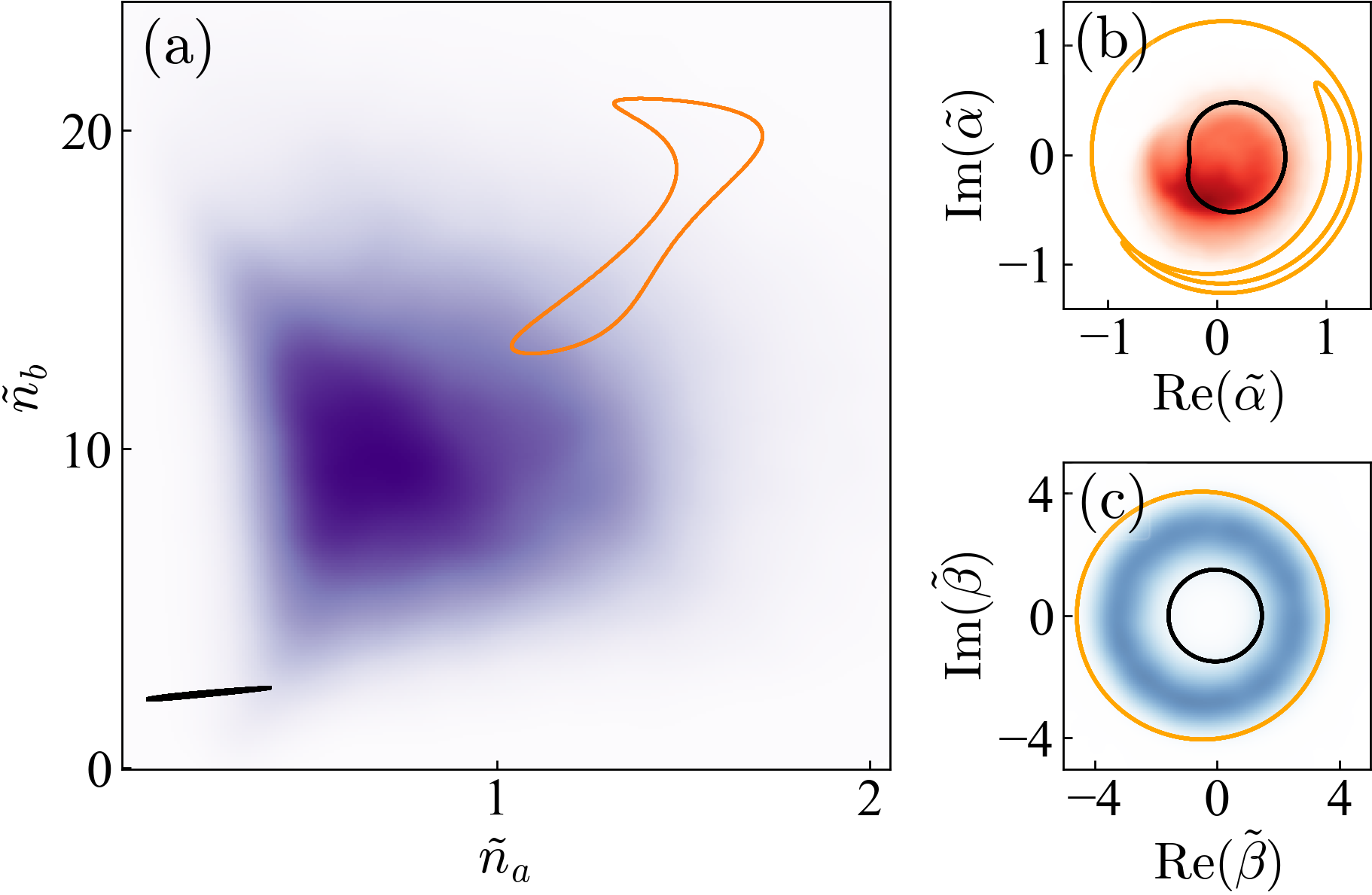}
	\caption{
		\textit{Quantum-melted regime.}
		Stationary phase-space density for $\aleph=1$, reconstructed from quantum-jump trajectory ensembles,
		(a) projected onto the $(\tilde n_a, \tilde n_b)$ population plane,
		and
		(b, c) projected onto the optical and mechanical complex phase space, respectively.
		Black and orange solid lines mark the two coexisting mean-field solutions \LC{1} and \LC{2}, respectively.
		In this regime, quantum fluctuations are comparable to the separation between attractors, enabling almost free movement between the basins of attraction. As a result, the separation of metastable states and their associated timescales is effectively melted, and metastability and directional switching can no longer be resolved.
	}
	\label{fig:app:quantum_stationary_state_aleph=1}
\end{figure}

\paragraph{Quantum-melted regime}
If quantum fluctuations are sufficiently strong, the bistable attractor structure is effectively washed out.
In Fig.~\ref{fig:app:quantum_stationary_state_aleph=1}, we show the stationary phase-space density for $\aleph=1$.
Panel (a) shows the distribution in the $(\tilde n_a, \tilde n_b)$ population plane, panels (b, c) show the density in the optical and mechanical complex phase space, respectively.
Rather than remaining localized within one of the two basins, the system exhibits rapid switching that dominates the dynamics. As a consequence, metastability and separation of timescales are no longer observed. In this quantum-melted regime, both the geometric organization of phase space and well-defined switching rates can no longer be extracted.
For the analysis in the main text, we therefore restrict to $\aleph\geq2$.
%

\section{Identification of metastable states using a Hidden Markov model}
\label{app:hmm}

In this Appendix, we detail the Hidden Markov model (HMM) procedure used to segment individual quantum-jump trajectories into metastable visits to the coexisting attractors \LC{1} and \LC{2}.
The HMM takes the trajectory-resolved observables as input and assigns, at each time step of each trajectory, a discrete latent state corresponding to one of the attractors.
The resulting state sequence forms the basis for the extraction of dwell-time distributions, survival functions, and directional escape rates, see Sec.~\ref{sec:traj_states}.
It further allows for the alignment of trajectory segments around switching events in order to reconstruct the typical transition dynamics and phase-space geometry of escape paths, see Sec.~\ref{sec:geom_escape}.

\subsection{Motivation}

Trajectory segmentation based on a direct reconstruction of the basins of attraction and the separatrix from the underlying quantum master equation would require resolving the full four-dimensional phase-space structure of the stochastic dynamics, which is computationally prohibitive for long quantum-jump trajectories.
Moreover, such a reconstruction relies on detailed knowledge of the dynamical model and therefore does not directly translate to experimentally recorded trajectories.
However, for the purposes of switching-rate extraction and transition analysis, a full geometric reconstruction is not necessary.

Instead, we exploit that experimentally accessible observables exhibit clearly distinct dynamical signatures when the system resides in the basin of \LC{1} or \LC{2}.
In the metastable regime, trajectory data cluster around two well-separated regions in observable space, reflecting localization near the corresponding deterministic limit cycles.
The time series therefore decomposes into two dynamical regimes, even though the underlying attractor label is not directly observable.

This setting is naturally captured by a Hidden Markov model (HMM)~\cite{Rabiner1986, Rabiner1989}.
Conceptually, an HMM assumes that an observed time series is generated by an underlying sequence of unobserved (\q{hidden}) states that evolve according to a Markov process, with each state producing characteristic statistical signatures in the measured observables.
This general concept appears across disciplines and makes HMMs a powerful tool for pattern recognition and sequence analysis~\cite{Rabiner1986,Rabiner1989}.
Here, the metastable attractors \LC{1} and \LC{2} are treated as the latent discrete states, while the measured trajectories of observables constitute continuous emissions.
The key physical assumption is a separation of timescales: intra-attractor fluctuations occur on short timescales, whereas noise-induced switching events are rare. Under this separation, the coarse-grained switching dynamics between the metastable states is well approximated by a Markov chain, while the emission model accounts for fluctuations within each attractor.
Fitting the HMM to the observation data and subsequent state decoding yields the most probable sequence of underlying states \LC{1}, \LC{2}.

This HMM approach to the trajectory-segmentation problem avoids heuristic thresholding in phase space or local window-based classification schemes, which typically require ad hoc parameter choices and are sensitive to transient excursions near the separatrix~\cite{Haenggi1990,Margiani2022,Sett2024}.
In contrast, the HMM performs a global probabilistic inference of the entire state sequence, combining information from the full trajectories and enforcing temporal consistency through the Markov prior.
Provided the two limit-cycle signatures remain sufficiently distinguishable in the chosen observables, the HMM framework produces a robust and globally consistent identification of residence intervals that can be directly used for dwell-time and switching-rate analysis.

\subsection{Model definition} 

\begin{figure}
	\centering
	\includegraphics[width=0.58\linewidth]{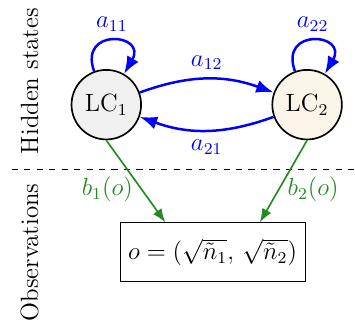}
	\caption{
		\textit{Hidden Markov model for trajectory segmentation.}
		Schematic representation of the two-state Hidden Markov model used to identify metastable dynamics. 
		The latent states correspond to the two limit cycles \LC{1} and \LC{2}, which evolve according to a Markov process with transition probabilities $a_{ij} = P(s_t = q_j \mid s_{t-1} = q_i)$. 
		Each state emits continuous observations $o_t = (\sqrt{\tilde{n}_a(t)}, \sqrt{\tilde{n}_b(t)})$ drawn from state-dependent emission distributions $b_i(o_t) = P(o_t \mid s_t = q_i)$, modeled as Gaussian densities. 
		The HMM provides a probabilistic mapping from observed trajectories to the underlying sequence of metastable states, forming the basis for dwell-time and switching-rate analysis.
	}
	\label{fig:app:hmm_setup}
\end{figure}

The input for training the HMM is a sequence of observations $o_t$, where $t=0,\ldots,T$ denotes the observation time.
The model assumes that these observations are generated by an underlying sequence of latent states $\{s_t\}$ that evolve according to a Markov process. Second, it assumes that the probability of an observation depends only on the current state of the system (\q{output independence}).
The HMM is built from the following components:
\begin{enumerate}[label=(\roman*)]
	\item A finite set of $N$ states
	\begin{equation}
		Q = \{ q_1, q_2, \ldots, q_N \} \;.
	\end{equation}
	These are the possible latent (\q{hidden}) states of the system.
	Each state $s_t$ in the state sequence $\{ s_t \}$ is an element of this set, $s_t\in Q$.
	The state sequence is assumed to evolve according to a first-order Markov process,
	\begin{equation}
		P(s_t \mid s_{t-1}, s_{t-2}, \ldots) = P(s_t \mid s_{t-1}) \;.
	\end{equation}
	\item A transition matrix
	\begin{equation}
		\boldsymbol{A} = (a_{ij}) \;, \;\;
		a_{ij} = P(s_t = q_j \mid s_{t-1} = q_i) \;,
	\end{equation}
	where $a_{ij}$ is the probability that the system transitions from state $q_i$ to state $q_j$, with $\sum_{j}a_{ij}=1$ for all $q_i \in Q$.
	\item Emission probability distributions
	\begin{equation}
		B = \{ b_i(o) \} \;,
	\end{equation}
	where
	\begin{equation}
		b_i(o_t) = P(o_t \mid s_t = q_i)
	\end{equation}
	denotes the probability of observing the emission $o_t$, given that the system is in state $s_t = q_i$.
	\item Initial state probabilities
	\begin{equation}
		\boldsymbol{\pi} = (\pi_i)\;, \;\;
		\pi_i = P(s_0 = q_i)\;, \;\;
		\sum_{i=1}^{N} \pi_i = 1 \;,
	\end{equation}
	used to initialize the Markov chain.
\end{enumerate}
Here, we consider $N=2$, with the latent states $q_1$, $q_2$ corresponding to the limit cycles \LC{1}, \LC{2}, respectively.
Each quantum-jump trajectory provides a sequence of bivariate observations
$o_t=(x_1(t),x_2(t))$,
where the features are defined in terms of the measured populations as
$x_1(t)=\sqrt{\langle n_a(t)\rangle}$ and $x_2(t)=\sqrt{\langle n_b(t)\rangle}$.
The emission distributions $B$ are modeled as multivariate Gaussian distributions in the two-dimensional observation space.
For each latent state $q_i$, the emission probability density is
\begin{equation}
	b_i(o) = \mathcal{N}(o \mid \boldsymbol{\mu}_i , \boldsymbol{\Sigma}_i) \;,
\end{equation}
with mean vector $\boldsymbol{\mu}_i$ and covariance matrix $\boldsymbol{\Sigma}_i$.
This provides an effective coarse-grained description of fluctuations around each attractor and does not assume Gaussian microscopic noise at the level of the underlying quantum dynamics.
The HMM setup of hidden states, observable features, and the emission model is visualized in Fig.~\ref{fig:app:hmm_setup}.

\subsection{Initialization}
The observation data $o_t$ are standardized using a global normalization fitted across the full ensemble of trajectories to ensure comparable scaling of the emission statistics.
The transition matrix is initialized with equal probabilities for all transitions,
\begin{equation}
	a_{ij} = \frac{1}{2} \;, \;\;\; i,j\in \{1,2\} \;,
\end{equation}
corresponding to a non-informative prior over state transitions.
The initial state distribution $\boldsymbol{\pi}$ is estimated empirically from the data as the relative frequency of states at the starting time of each trajectory.
The parameters of the emission distributions are initialized using a clustering-based procedure. Specifically, the concatenated observation data from all standardized trajectories are first partitioned into $N=2$ clusters using the K-means clustering algorithm. The resulting cluster centers and variances are used as initial estimates $(\boldsymbol{\mu}_i,\boldsymbol{\Sigma}_i)$ of the Gaussian emission distributions.

\subsection{Training and State Decoding}
\begin{figure}
	\centering
	\includegraphics[width=1\linewidth]{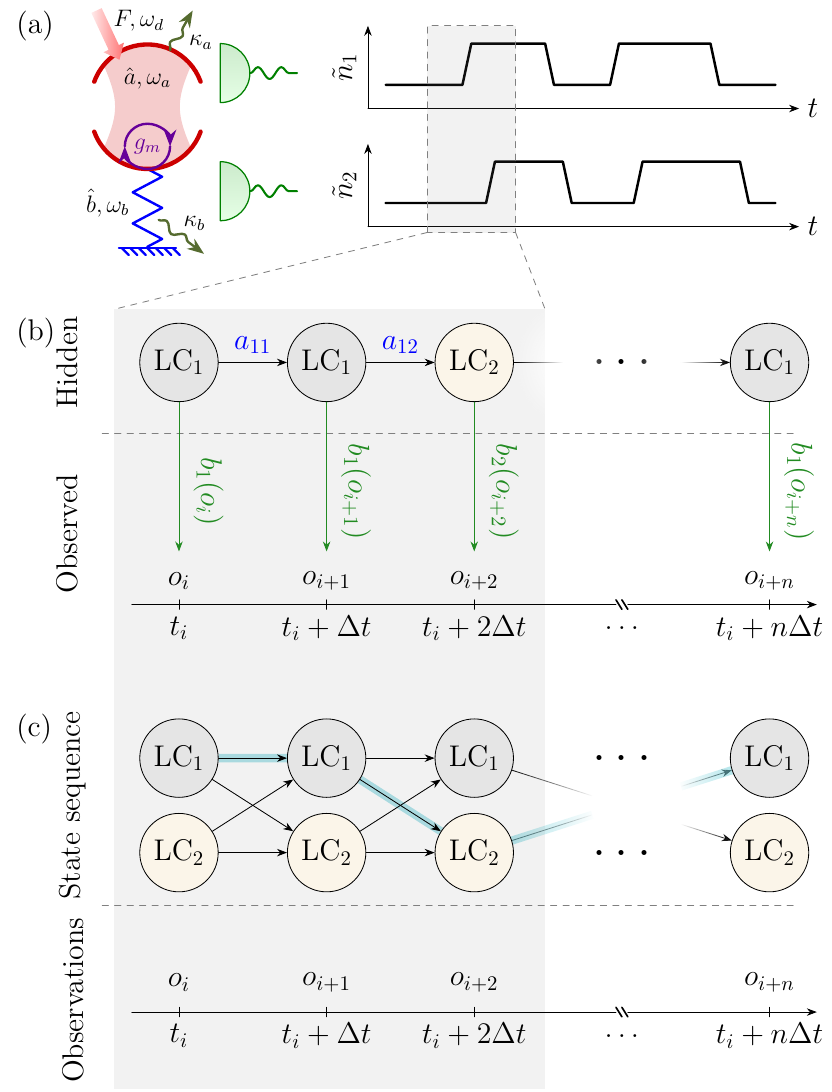}
	\caption{
		\textit{Hidden Markov model for quantum trajectory segmentation.}
		(a) Schematic of the measured driven-dissipative optomechanical system leading to time traces of the measured observables $\tilde{n}_1(t)$ and $\tilde{n}_2(t)$, exhibiting switching between metastable dynamical regimes. The shaded window indicates the zoomed-in region of the time trace for which the discrete time steps of the HMM analysis are illustrated below.
		(b) Graphical representation of the time-discrete state and emission sequences in the Hidden Markov model. Each measured observation $o_t$ is produced by a latent hidden state (here, limit cycles \LC{1} and \LC{2}). The states evolve according to a discrete-time Markov chain with transition probabilities $a_{ij}$ and emit observable data according to the state-dependent distributions $b_i(o)$.
		Model parameters are iteratively fitted on the entire observed trajectories using the Baum-Welch algorithm.
		(c) Decoding of the most probable sequence of hidden states from the observed trajectory using the Viterbi algorithm. The inferred state sequence is used to segment the trajectory into metastable visits to \LC{1} and \LC{2}, forming the basis for dwell-time statistics and switching-rate extraction.
	}
	\label{fig:app:hmm_2}
\end{figure}

The HMM parameters, including transition probabilities and emission statistics, are learned from the trajectory data using maximum-likelihood estimation via the Expectation-Maximization (EM) algorithm for Hidden Markov models, also known as Baum-Welch algorithm~\cite{Baum1970, Baum1972}.
The training is performed jointly across all trajectories by maximizing the summed log-likelihood of all sequences.
After training, the most probable sequence of hidden states for each trajectory is inferred using the Viterbi algorithm~\cite{Viterbi1967,Forney1973}, which computes the maximum-likelihood state path consistent with the learned model parameters.
A schematic representation of the time-discrete state and emission sequences in the HMM is visualized in Fig.~\ref{fig:app:hmm_2}.

The inferred state sequences form the basis for the extraction of dwell times and switching statistics used in the main text.
We verified that the inferred residence times are robust with respect to moderate variations in the initialization of the model parameters and the covariance parameterization of the emission distributions.

\subsection{Scope and limitations}
\label{app:hmm_limitations}

The Hidden Markov model employed here provides an effective coarse-grained description of the metastable switching dynamics between the two limit-cycle attractors. Its validity relies on several assumptions that define the scope of applicability and impose limitations on the interpretation of the results.
First, the HMM assumes that the underlying dynamics of the latent states is Markovian, i.e., that the transition probability between attractors depends only on the current state and not on the past trajectory.
In practice, this assumption should be understood as an effective description at the level of coarse-grained metastable states. The survival functions observed here exhibit deviations from single-exponential behavior at short times, indicating memory effects due to rapid recrossings at the basin boundary and unresolved transient intra-attractor dynamics, which are not captured by the Markov approximation. However, at longer times the survival functions converge to a single-exponential decay. This demonstrates the existence of a dominant slow timescale governing the transitions between the attractors. The Markovian assumption is therefore justified as an effective description of the long-time switching dynamics, even in the absence of a strict separation of timescales.
Second, the performance of the HMM relies on a sufficient separation of the emission distributions associated with the two attractors in the chosen observable space. In our implementation, the features $o_t = (\sqrt{\langle n_a(t) \rangle}, \sqrt{\langle n_b(t) \rangle})$ form well-separated clusters in the metastable regime, enabling robust state identification. If the distributions overlap significantly, for instance in the regime of strong quantum fluctuations where the attractor structure becomes blurred, see Fig.~\ref{fig:app:quantum_stationary_state_aleph=1} in \appref{app:Wigner_stationary_state_and_quantum_melted_regime}, the inferred state sequence becomes ambiguous and the extracted switching statistics lose reliability.
Third, the Gaussian emission model provides an effective description of fluctuations around each attractor but does not capture non-Gaussian features of the underlying quantum dynamics. In particular, correlations along the limit cycle and phase-dependent fluctuations are averaged over in this coarse-grained representation.
Finally, the HMM yields a discrete two-state representation of the dynamics and therefore cannot resolve internal structure within each attractor. In particular, it does not distinguish between different phases along the limit cycle or between distinct escape pathways originating from different regions of phase space. As a consequence, the HMM-based segmentation is well suited for extracting long-time switching kinetics and dwell-time statistics, but not for resolving the detailed geometry of escape trajectories or phase-resolved activation processes, which require complementary analysis as presented in Sec.~\ref{sec:geom_escape}.
Within these limitations, the HMM provides a robust and statistically consistent framework for extracting metastable residence times and directional switching rates from long stochastic trajectories (from numerical or experimental data).

\section{Survival analysis and extraction of the switching rates}
\label{app:survival_analysis}


\begin{figure}[]
	\includegraphics[width=0.95\linewidth]{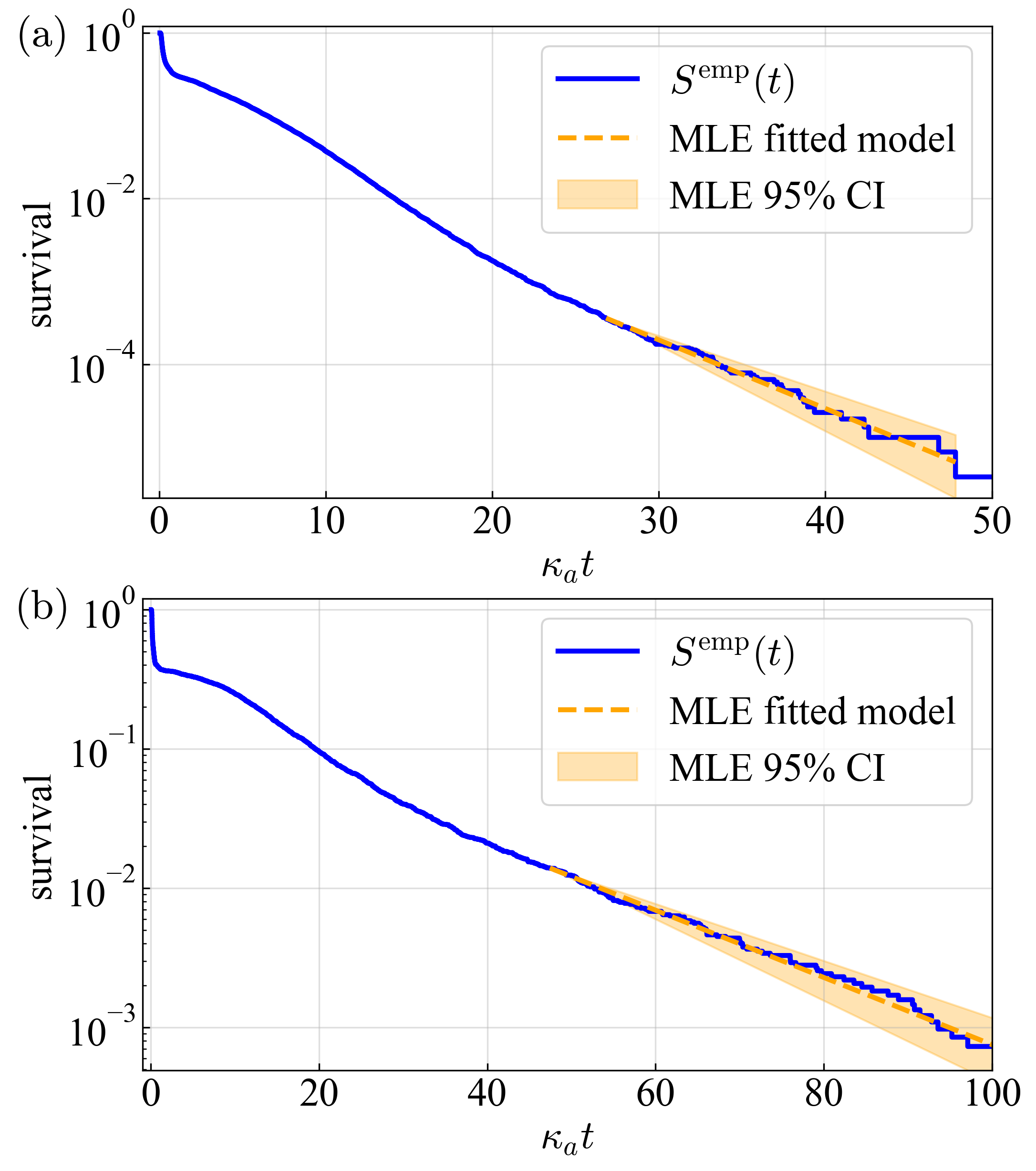}
	\caption{
		\textit{Empirical survival functions and conditional survival model fitted to long-time dwell-time statistics for metastable dwell times in \LC{1}}.
		The survival probability $S^{\mathrm{emp}}(t)$ (blue solid line), estimated via the Kaplan-Meier estimator, is shown for two values of the scaling parameter $\aleph$. 
		At short times, $S^{\mathrm{emp}}(t)$ exhibits a rapid initial decay due to transient recrossings (“flickering”) near the basin boundary and incomplete relaxation into the attractor. 
		At intermediate times, deviations from a single exponential reflect relaxation dynamics within the basin. 
		At long times, the survival function develops a clear exponential tail, indicating Markovian switching governed by a single dominant escape rate.
		The dashed orange line shows the maximum-likelihood fit to the exponential conditional survival model in Eq.~\ref{eq:app:conditional_survival_model},  which is restricted to dwell times $T_{\mathrm{dwell}}>t_0$, with the shaded region indicating the corresponding $95\%$ confidence interval obtained via bootstrap resampling. 
		The good agreement in the asymptotic regime validates the extraction of the directional switching rate $k_{12}$ used in the main text.
	}
	\label{fig:app:survival_LC1}
\end{figure}

%
\begin{figure}[]
	\includegraphics[width=0.95\linewidth]{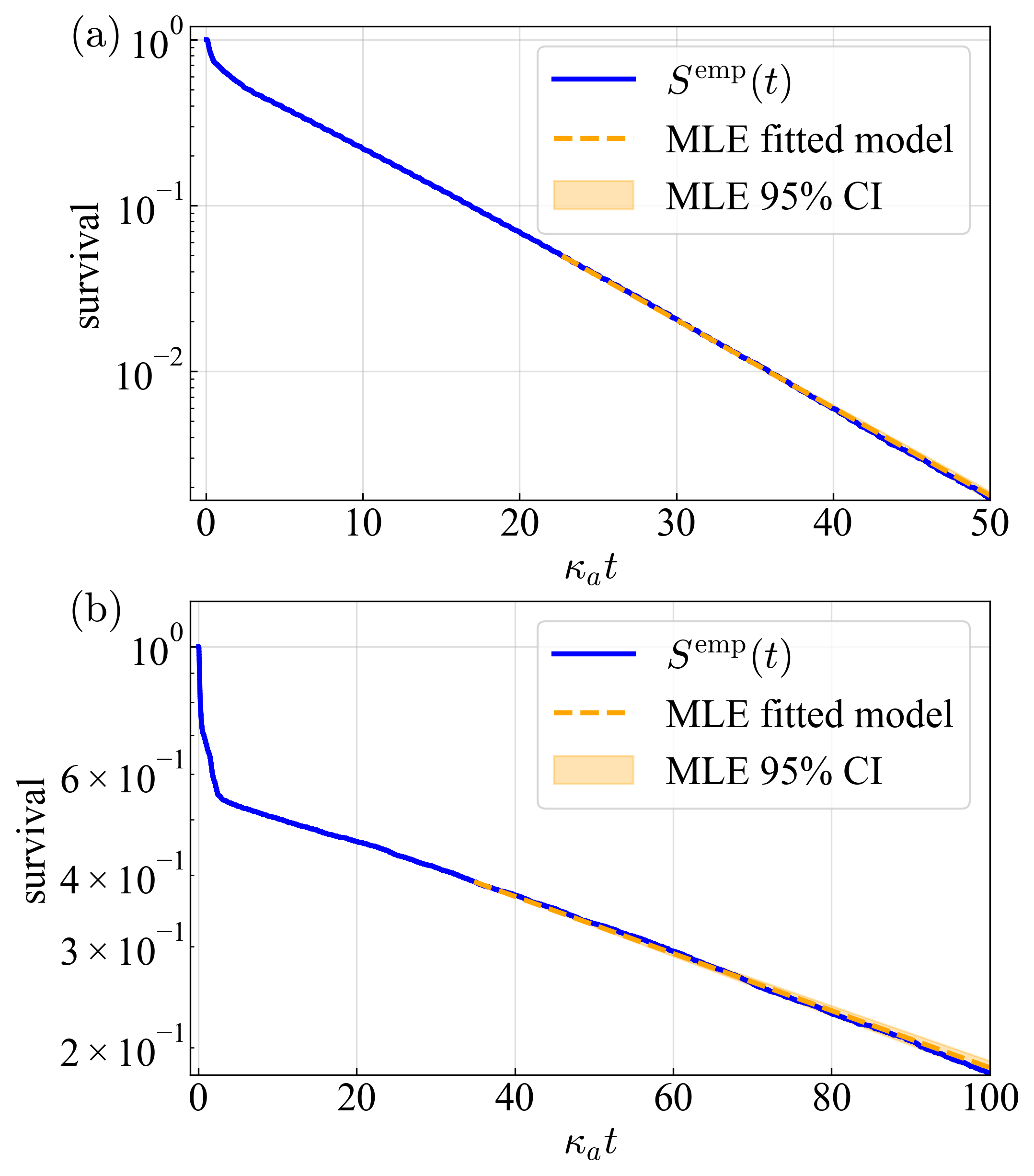}
	\caption{
		\textit{Empirical survival functions and conditional survival model fitted to long-time dwell-time statistics for metastable dwell times in \LC{2}.}
		Same as Fig.~\ref{fig:app:survival_LC1}, but for dwell times in \LC{2}. 
		Compared to \LC{1}, the survival functions exhibit a shorter intermediate-time regime before reaching the exponential tail, reflecting faster relaxation dynamics within the basin after switching.
		At long times, the survival functions again approach a clear exponential decay, validating the extraction of the asymptotic switching rate $k_{21}$ used in the main text.
	}
	\label{fig:app:survival_LC2}
\end{figure}

%

If the system switches the basins of \LC{1} and \LC{2}, each visit to one of the basins can be characterized by its dwell time $T_{\text{dwell}}$, given by the duration between the entry into the state and the subsequent transition to the opposite state.
Here, we analyze the dwell-time statistics of both \LC{1} and \LC{2} and detail how the directional switching rates discussed in Sec.~\ref{sec:traj_states} are extracted.

First, we consider the survival function
\begin{equation}
	S_i(t)= P(T_{\text{dwell}} > t\mid\text{entry into state }\LC{i})\;,
	\label{eq:app:survival_function}
\end{equation}
which measures the probability that a trajectory that has entered \LC{i} persists longer than time $t$ within this basin, without switching to the other basin.
%
If the exact survival function of a system is not known, it needs to be estimated based on empirical data. Here, we use the Kaplan-Meier estimator to compute the empirical survival function $S^{\text{emp}}_i(t)$, which provides a nonparametric estimator of the true underlying survival function $S_i(t)$ \cite{Kaplan1958}.
We extract the dwell times from the quantum-jump trajectories, which are segmented into metastable visits to the two basins by the hidden Markov model as described in Sec.~\ref{app:hmm}.
Since the quantum-jump trajectories have finite length, the dwell times can be biased if a segment is truncated by the boundaries of the observation interval, which is referred to as left- and right-censoring.
In the dwell time statistics, we only include incident runs, i.e., visits where the basin was entered during the observation interval, thus filtering out left-censored data.
The right-censoring is automatically treated correctly by the Kaplan-Meier estimator
\begin{equation}
	S^{\text{emp}}_i(t) = \prod_{t_{(k)} \le t} \left(1 - \frac{d_k}{n_k}\right)\;,
	\label{eq:app:KM_estimator}
\end{equation}
where $d_k$ denotes the number of events happening at time $t_{(k)}$, and $n_k$ is the number of objects at risk at time $t_{(k)}$.
Here, an \q{event} corresponds to the trajectory leaving the basin \LC{i}, and the \q{objects at risk} are the trajectories that still remain in the basin.

In Fig.~\ref{fig:app:survival_LC1} and Fig.~\ref{fig:app:survival_LC2}, the blue solid lines display the empirical survival functions $S^{\text{emp}}_i(t)$ for \LC{1} and \LC{2}, respectively, for two values of the scaling parameter $\aleph$.
For both \LC{1} and \LC{2}, we observe an initial rapid drop of the survival probability.
This is a consequence of the \q{flickering} events at the basin boundary, which occur on a timescale comparable to the inverse dissipation constant $1/\kappa_a$.
The initial drop is followed by an intermediate regime, where the system has not yet fully relaxed into the attractor. 
Deviations from a single exponential in this regime reflect intra-attractor relaxation dynamics and transient recrossings near the basin boundary, which are not resolved in the effective two-state description.
After relaxation, the asymptotic behavior of the empirical survival functions exhibits an exponential tail, consistent with Markovian switching dynamics and activated escape governed by a single asymptotic rate $k_{ij}$.


Motivated by the observed exponential long-time behavior of the empirical survival functions, we assume an exponential survival model
\begin{equation}
	S_i(t) = e^{- k_{ij} t}\;,
\end{equation}
corresponding to a constant escape rate $k_{ij}$, and restrict the model to dwell times satisfying $T_{\text{dwell}}> t_0$, where $t_0$ marks the onset of the asymptotic in the empirical survival function. This leads to the conditional survival model
\begin{equation}
	S_i(t \mid T_{\text{dwell}}> t_0) = \frac{S_i(t)}{S_i(t_0)} = e^{-k_{ij}(t-t_0)} \; .
	\label{eq:app:conditional_survival_model}
\end{equation}
We perform a maximum-likelihood estimation (MLE) to this parametric model to extract the directional switching rates $k_{ij}$. This is carried out directly on the dwell-time distribution obtained from the segmented trajectories, restricting to incident segments.
The MLE accounts for right-censoring by including censored dwell times through their contribution to the survival probability, while completed segments contribute via the corresponding probability density.
Uncertainties of the fitted rates $k_{ij}$ are estimated via bootstrap resampling of the dwell-time data, with confidence intervals obtained from the empirical distribution of fitted parameters.
The fitted survival models are shown in Fig.~\ref{fig:app:survival_LC1} and Fig.~\ref{fig:app:survival_LC2}, demonstrating good agreement with the nonparametric Kaplan-Meier estimates $S^{\text{emp}}(t)$.
The resulting rates $k_{ij}$ are presented in Fig.~\ref{fig: rates_vs_aleph} in the main text.

\section{Phase-conditioned escape hazard}
\label{app:hazard}

As explained in Section~\ref{sec:geom_escape}, the event-conditioned phase distributions $P(\phi_a,\tau \mid \LC{i}\to\LC{j})$ shown in Fig.~\ref{fig: phase_hazard} reflect the conditional probability that the system is at phase $\phi_a$ at time $\tau$ relative to a switching event \LC{i} $\to$ \LC{j} at $\tau=0$.
These distributions are biased by the stationary phase distributions within the basins: If the system spends more time at a specific phase, it will more likely escape from there, even if the associated action costs might be larger than for escape channels starting at other phases along the cycle.
The distributions shown in Fig.~\ref{fig: phase_hazard} therefore do not directly represent the intrinsic, phase-dependent escape propensity or the action costs associated with different escape paths.

Here, we present the event-conditioned phase distributions normalized by the stationary distributions as
\begin{equation}
	P(\tau\mid\phi_a,\LC{i}\to\LC{j}) = \frac{P(\phi_a,\tau\mid\LC{i}\to\LC{j})}{P_{\text{stat}}(\phi_a\mid\text{LC}(\tau))}\;,
\end{equation}
where $P_{\text{stat}}(\phi_a \mid \LC{i})$ and $P_{\text{stat}}(\phi_a \mid \LC{j})$ denote the stationary phase distributions within the basins of \LC{i} and \LC{j}, respectively, and
\begin{align}
	\text{LC}(\tau) =
	\begin{cases}
		\LC{i} & \text{for } \tau \leq 0 \\
		\LC{j} & \text{for } \tau > 0
	\end{cases} \;.
\end{align}
This normalization therefore removes the bias from the nonuniform stationary phase occupancy along the oscillatory state, thereby isolating the intrinsic phase dependence of escape.

In the limit $\tau \to 0^-$, the normalized distribution can be interpreted as a phase-conditioned escape hazard,
\begin{equation}
	h_i(\phi_a)\equiv P(\tau\to 0^- \mid \phi_a, \LC{i}\to\LC{j}) \;,
\end{equation}
measuring the relative \q{risk} of escape at each phase, analogous to a hazard rate in survival analysis, but defined in a phase- and event-conditioned sense rather than as a time-dependent rate.
It is thus naturally interpreted in terms of the phase-dependent activation cost $S(\phi_a)$, with maxima corresponding to lower-action escape pathways, while suppressed regions reflecting higher activation cost.
For the discrete numerical timesteps of the trajectories, we estimate this effective hazard directly from the distribution at the switching time $\tau=0$.

\begin{figure}[H]
	\centering
	\includegraphics[width=1\linewidth]{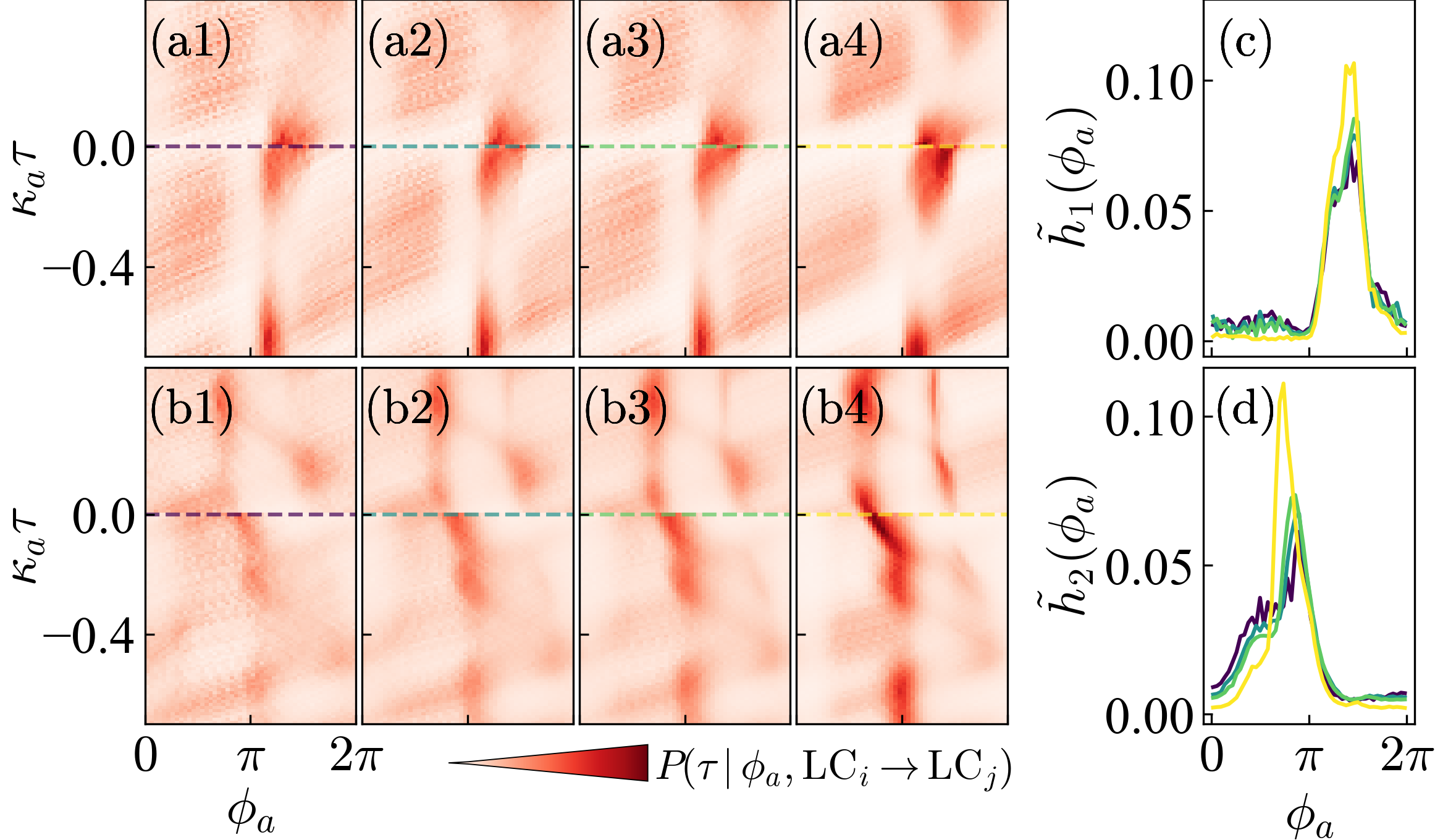}
	\caption{
		\textit{Event-conditioned phase distributions normalized by the stationary phase distributions, and phase-conditioned escape hazard.}
		(a,b) Normalized event-conditioned phase distributions $P(\tau\mid \phi_a,\LC{i}\to\LC{j}$), obtained by normalizing the event-conditioned phase distributions by the stationary phase occupancy, for (a) \LC{1}$\to$\LC{2} and (b) \LC{2}$\to$\LC{1}.
		Columns~$1-4$ correspond to increasing fluctuations strengths, $\aleph=2.5$, $\aleph=3.0$, $\aleph=3.5$, and $\aleph=9.0$.
		The normalization removes the bias due to the nonuniform stationary phase distribution and isolates the intrinsic phase-dependent escape propensity.
		(c,d) Phase-conditioned escape hazard $\tilde{h}_i(\phi_a)$, corresponding to the normalized event-conditioned phase distributions at the boundary crossing at $\tau\to 0^-$, rescaled to unity.
		For \LC{1}$\to$\LC{2}, the hazard is sharply localized, indicating a single dominant escape channel.
		For \LC{2}$\to$\LC{1}, multiple peaks are visible, reflecting competing phase-dependent escape pathways, with peak heights encoding their relative activation costs.
	}
	\label{fig:app:hazard}
\end{figure}

Figure~\ref{fig:app:hazard} shows the normalized event-conditioned phase distributions $P(\tau\mid \phi_a,\LC{i}\to\LC{j})$ for (a1-a4) switching events $\LC{1}\to\LC{2}$ and (b1-b4) switching events $\LC{2}\to\LC{1}$, where the columns correspond to increasing values of the scaling parameter $\aleph$.
The colored dashed lines highlight the time $\tau=0$, at which the system crosses the basin boundary.
Panels (c),(d) display the corresponding phase-conditioned escape hazard $\tilde h_i(\phi_a)$, obtained from the normalized distributions in the limit $\tau\to 0^-$, rescaled to unity.
The figure is thus analogous to Fig.~\ref{fig: phase_hazard} in the main text, but with the event-conditioned phase distributions normalized by the stationary phase distributions, thereby removing the bias due to nonuniform stationary phase occupancy along the oscillatory state and isolating the intrinsic phase dependence of escape.
As in Fig.~\ref{fig: phase_hazard}, the two switching directions exhibit a pronounced qualitative difference.
For $\LC{1}\to\LC{2}$, the risk of escape is strongly localized within a narrow phase interval and remains almost independent of the fluctuation strength across the observed values, consistent with a single dominant escape corridor which has a significantly lower activation cost than competing pathways.
In contrast, for $\LC{2}\to\LC{1}$, multiple distinct phase regions show significant risk of escape, with the relative weights of these peaks shifting as the fluctuation strength increases. Subdominant pathways become more prominent at stronger fluctuations (small $\aleph$), consistent with the crossover behavior in the switching-rate scaling discussed in Sec.~\ref{sec:action_based_interpretation}, and demonstrating the coexistence of several competing escape channels.

A direct comparison with Fig.~\ref{fig: phase_hazard} highlights the role of the normalization.
In particular, for $\LC{2}\to\LC{1}$ the raw event-conditioned phase distribution remains sizable in the interval $\phi_a\in[\pi,2\pi]$, whereas the normalized hazard is strongly suppressed in this region.
This shows that escape from these phases is intrinsically less favorable (higher activation cost), but occurs frequently in the unnormalized distribution due to the increased time spent at these phases along the limit cycle.

Overall, the phase-conditioned hazard provides direct evidence that the multichannel structure inferred from the switching-rate scaling is intrinsic to the escape dynamics, consistent with a phase-dependent activation landscape, and cannot be explained solely by variations in the stationary phase distribution.

\section{Liouvillian gap and metastable switching}
\label{app:liouvillian_gap}

In this section, we relate the timescales of the switching dynamics to the Liouvillian spectrum, showing how the metastability is reflected by the slow relaxation mode.
The dynamics of the open quantum system is governed by the Liouvillian superoperator $\mathcal L$ via
\begin{equation}
	\partial_t \hat\rho
	=
	\mathcal L \hat\rho\;.
\end{equation}
Its spectrum $\{\lambda_\alpha\}$ is complex, with stationary states corresponding to zero eigenvalues and $\mathrm{Re}(\lambda_\alpha)<0$ determining decay rates~\cite{Minganti2018}.
If the leading nonzero Liouvillian eigenvalue $\lambda_1$ is isolated, the asymptotic relaxation rate toward the stationary subspace is determined by the Liouvillian gap
\begin{equation}
	\Lambda
	=
	-\mathrm{Re}(\lambda_1)\;.
\end{equation}
In the metastable regime considered here, a clear separation between fast intra-attractor relaxation and rare switching defines a reduced slow manifold spanned by two long-lived dynamical states~\cite{Macieszczak2016,Rose2016,Macieszczak2021} corresponding to \LC{1} and \LC{2}. 
Projecting onto their occupations $p_i(t)$ yields an effective two-state Markov dynamics,
\begin{equation}
	\partial_t
	\begin{pmatrix}
		p_1\\
		p_2
	\end{pmatrix}
	=
	\begin{pmatrix}
		-k_{12} & k_{21}\\
		k_{12} & -k_{21}
	\end{pmatrix}
	\begin{pmatrix}
		p_1\\
		p_2
	\end{pmatrix}\;,
\end{equation}
where $k_{ij}$ are the directional switching rates introduced in the main text.
The corresponding eigenvalues are the stationary eigenvalue $\lambda_0^{\mathrm{eff}}=0$ and the relaxation eigenvalue
\begin{equation}
	\lambda_1^{\mathrm{eff}}
	=
	-(k_{12}+k_{21})\;,
\end{equation}
showing that the slow Liouvillian relaxation $\Lambda_{\mathrm{eff}} = -\lambda_1^{\mathrm{eff}}$
is directly governed by the directional switching dynamics. 
The effective Liouvillian relaxation rate therefore inherits the multichannel structure of the underlying escape dynamics through
\begin{equation}
	\Lambda_{\mathrm{eff}}(\aleph)
	=
	k_{12}(\aleph)+k_{21}(\aleph)\;.
\end{equation}
The individual directional rates $k_{ij}$, however, cannot be independently read off from this Liouvillian eigenvalue; rather, the eigenvalue only fixes their sum, while their ratio determines the stationary state.

\bibliographystyle{KilianStyle}
\bibliography{biblio_twin_LC}
	
\end{document}